\documentclass[12pt]{article}
\usepackage{amssymb}
\usepackage{amsmath}
\usepackage{amsfonts}
\usepackage{mitpress}

\usepackage[T1]{fontenc}
\usepackage[latin1]{inputenc}
\usepackage{bbm}
\usepackage{stmaryrd}
\usepackage{latexsym}
\usepackage{amscd}
\usepackage{yfonts}
\usepackage{dsfont}
\usepackage{mathabx}
\usepackage[dvips]{graphicx}
\usepackage{epsfig}
\usepackage[font={small}]{caption}
\usepackage{color}
\usepackage{float}
\usepackage{upgreek}
\usepackage{tikz}
\usepackage{tikz-cd}
\usepackage{relsize}
\usepackage{url}

\DeclareFontFamily{U}{txsyc}{}
\DeclareFontShape{U}{txsyc}{m}{n}{
   <-> txsyc%
}{}
\DeclareFontShape{U}{txsyc}{bx}{n}{
   <-> txbsyc%
}{}
\DeclareFontShape{U}{txsyc}{l}{n}{<->ssub * txsyc/m/n}{}
\DeclareFontShape{U}{txsyc}{b}{n}{<->ssub * txsyc/bx/n}{}
\DeclareSymbolFont{symbolsC}{U}{txsyc}{m}{n}
\SetSymbolFont{symbolsC}{bold}{U}{txsyc}{bx}{n}
\DeclareFontSubstitution{U}{txsyc}{m}{n}
\DeclareMathSymbol{\df}{\mathrel}{symbolsC}{"42}
\DeclareMathSymbol{\fd}{\mathrel}{symbolsC}{"43}
\DeclareMathSymbol{\lJoin}{\mathrel}{symbolsC}{"58}
\DeclareMathSymbol{\rJoin}{\mathrel}{symbolsC}{"59}

\newcommand{\f}[2]{\frac{#1}{#2}}
\newcommand{\cA}{\mathcal{A}}
\newcommand{\cB}{\mathcal{B}}
\newcommand{\cC}{\mathcal{C}}
\newcommand{\cD}{\mathcal{D}}

\newcommand{\cF}{\mathcal{F}}
\newcommand{\cG}{\mathcal{G}}

\newcommand{\cJ}{\mathcal{J}}
\newcommand{\cK}{\mathcal{K}}
\newcommand{\cL}{\mathcal{L}}

\newcommand{\cM}{\mathcal{M}}

\newcommand{\NN}{\mathbb{N}}

\newcommand{\RR}{\mathbb{R}}

\newcommand{\iy}{\infty}

\newcommand{\lt}{\left}
\newcommand{\me}{\medskip}

\newcommand{\pa}{\partial}
\newcommand{\ri}{\rightarrow}
\newcommand{\rt}{\right}
\newcommand{\sm}{\smallskip}
\newcommand{\tr}{\triangle}

\newcommand{\wi}{\widetilde}

\newcommand{\fo}{\forall\ }

\newcommand{\lve}{\lt\vert}
\newcommand{\lVe}{\lt\Vert}

\newcommand{\rve}{\rt\vert}
\newcommand{\rVe}{\rt\Vert}

\newcommand{\st}{\,:\,}

\newcommand{\un}{\mathds{1}}

\newcommand{\bq}{\begin{eqnarray*}}
\newcommand{\bqn}[1]{\begin{eqnarray}\label{#1}}
\newcommand{\eq}{\end{eqnarray*}}
\newcommand{\eqn}{\end{eqnarray}}
\newcommand{\wwtbp}{\par\hfill $\blacksquare$\par\me\noindent}

\newcommand{\ttsim}{\raise.17ex\hbox{$\scriptstyle\mathtt{\sim}$}}

\newcommand{\kh}{\kern .08em}

\newtheorem{pro}{Proposition} 
\newtheorem{cor}{Corollary}
\newtheorem{lem}{Lemma}

\renewcommand{\thepro}{\arabic{pro}}

\newenvironment{rem}
{\par\me\refstepcounter{pro}\noindent{\bf Remark \thepro\ }}
{\par\hfill $\square$\par\sm\noindent}

\newcommand{\prooff}{\par\me\noindent\textbf{Proof}\par\sm\noindent}
\newcommand{\proofff}[1]{\par\me\noindent\textbf{#1}\par\sm\noindent}

\newcommand{\rVtv}{\rVe_{\mathrm{tv}}}
\newcommand{\mus}{{\mu^*}}
\newcommand{\musr}{{\mu^*_r}}
\newcommand{\muss}{{\mu^*_{\mathrm{s}}}}
\newcommand{\musc}{{\mu^*_{\mathrm{c}}}}
\newcommand{\musa}{{\mu^*_{\mathrm{a}}}}
\newcommand{\z}{v}

\newtheorem{theorem}{Theorem}

\newdimen \dummy
\dummy=\oddsidemargin
 \input{tcilatex}
\begin{document}

\title{Optimal epidemic suppression under an ICU constraint\thanks{%
We wish to thank Tommy Andersson, Hannes Malmberg and Robert \"{O}stling for
valuable comments.}}
\author{Laurent Miclo\thanks{%
Toulouse Institute of Mathematics, Toulouse School of Economics and CNRS. Email:
laurent.miclo@math.cnrs.fr.}, Daniel Spiro\thanks{%
Department of Economics, Uppsala University.     Email:
daniel.spiro.ec@gmail.com.} \ and J\"{o}rgen Weibull\thanks{%
Department of Economics, Stockholm School of Economics.  Email: \mbox{jorgen.weibull@hhs.se}.}}
\maketitle
\date{\today}

\begin{abstract}
How much and when should we limit economic and social activity to ensure
that the health-care system is not overwhelmed during an epidemic? We study
a setting where ICU resources are constrained while suppression is costly
(e.g., limiting economic interaction). Providing a fully analytical solution
we show that the common wisdom of \textquotedblleft flattening the
curve\textquotedblright , where suppression measures are continuously taken
to hold down the spread throughout the epidemic, is suboptimal. Instead, the
optimal suppression is discountinuous. The epidemic should be left
unregulated in a first phase and when the ICU constraint is approaching
society should quickly lock down (a discontinuity). After the lockdown
regulation should gradually be lifted, holding the rate of infected constant
thus respecting the ICU resources while not unnecessarily limiting economic
activity. In a final phase, regulation is lifted. We call this strategy
\textquotedblleft filling the box\textquotedblright .

\pagebreak
\end{abstract}

\section{Introduction}

Amid the Covid-19 health and economic crisis one question stood at the
centre of professional opinion: How much and when should we limit economic
and social activity to ensure that the health-care system is not
overwhelmed? This question embodies two simultaneous goals when fighting a
pandemic: (1) To ensure that each infected person gets the best possible
care, we need to ensure that the capacity of the health-care system
(henceforth the ICU constraint) is never breached. Under Covid-19 the ICU
constraint is essentially the number of available respirators, indeed a
scarce resource in most countries. It was perhaps best epitomized by the UK
slogan \textquotedblleft Protect the NHS\textquotedblright \ and by the
Empirical College report (Ferguson et al., 2020). (2) The more one is
suppressing the spread the costlier it is since, absent a vaccine,
suppression boils down to keeping people away from each other thus limiting
economic and social life.

This paper extends the standard S.I.R. model (Kermack and McKendrick, 1927)
with those two extensions to provide an \textit{analytical} answer to the
above question. Our answer departs from the common wisdom. During the
Covid-19 pandemic, authorities, news reporting and policy makers popularized
the ideal policy as \textquotedblleft flattening the curve\textquotedblright
,\footnote{%
See, e.g., the Empirical College report (Ferguson et al., 2020), Branswell
(2020), Time (2020), Pueyo (2020),\ even Donald Trump (The Sun, 2020) and
many more.} i.e., imposing continuous limitations to lower the number of
simultaneously infected in all time periods. This would ensure that the peak
of the curve never crossed the ICU constraint. We show that this policy is
suboptimal. Instead, the optimal policy is characterized by what we call
\textquotedblleft filling the box\textquotedblright \ and a discontinuous
suppression. More precisely, it prescribes (Theorem \ref{Thm: Main} and
Figure \ref{Fig: Time dynamics}) leaving the spread unregulated during a
first phase. As the number of infected approaches the ICU constraint we
enter a second phase where harsh suppression measures are imposed at once (a
discontinuity) but afterwards gradually relaxed. The aim of policy in this
second phase is to stop the number of infected just below the ICU constraint
and keep it constant at that level. The discontinuous tightening followed by
gradual relaxation of suppression is optimal since the underlying growth of
infections is highest in the beginning of this phase. In a third phase, once
the underlying growth of infections subsides, no suppression measures are
taken.

The logic behind this result is simple, but bears relevance for a disease
spreading such as Covid-19. When access to a vaccine is not realistic within
a sufficiently near future and pinpointing each infectious person is not
feasible, which is implicitly assumed in our model, full eradication is not
possible. What remains then, is letting the infection spread in the
population but ensuring that each person gets best possible care, i.e.,
ensuring the available respirators are sufficient at all times. But there is
no point in leaving some of the respirators idle (if considering risk one
can view the ICU constraint as being the number of respirators with a
margin). Hence, early suppression is unnecessary and costly. Once the number
of infections reaches the ICU constraint, drastic suppression has to be
installed to keep it below. But also here it is unnecessarily costly to
suppress the whole curve as it leaves idle respirators. Hence, the aim
during the second phase is to precisely fill the ICU capacity. Once the
infection rate goes down so that the ICU constraint is no longer binding --
the third phase -- suppression can be lifted. The number of respirators and
the time axis can essentially be thought of as a box. \textquotedblleft
Filling the box\textquotedblright \ then simply means respecting the ICU
constraint while not incurring costs to leave idle resources. In the
concluding remarks we further discuss how various implicit assumptions may
change this result.

Apart from the policy implication our main contribution lies in the model
and the analysis itself. We develop and show how to fully analytically solve
an epidemic-economic model for the optimal suppression policy. Importantly,
the suppression policy is allowed to be fully time varying. Our approach is
thus clearly distinguished from a large number of recent papers (not least in economics)
that analyze policies numerically (e.g., Wearing et al., 2005; Iacoviello
and Liuzzi, 2008; Lee et al., 2011; Kar and Batabyal, 2011; Iacoviello and
Stasio, 2013; Giamberardino and Iacoviello, 2017; Gollier, 2020; Wang, 2020;
Farboodi et al, 2020; Eichenbaum et al., 2020; Alvarez et al, 2020). In
order to make analytical headway we abstract from many nuances that such,
numerical, papers consider including the possibility of testing (Gollier,
2020; Wang, 2020, Berger et al., 2020), the arrival of a vaccine (Zaman et
al., 2008; Iacoviello and Liuzzi, 2008; Lee et al., 2011; Kar and Batabyal,
2011; Giamberardino and Iacoviello, 2017; Farboodi et al, 2020), treatment
and education (e.g., Bakare et al., 2014) group heterogeneity (e.g., Shim,
2013;\ Sj\"{o}din, 2020), contact tracing (see, e.g., Wearing et al., 2005;
McCaw and McVernon, 2007; Britton and Malmberg, 2020; and references
therein), time delays (Zaman et al., 2009), network effects (e.g., Gourdin,
2011) and individual decision making (Farboodi et al, 2020;\ Eichebaum et
al., 2020). Our exercise is a stepping stone for considering also such
aspects in future work. To our knowledge ours it the first paper to at all
consider optimal policy under an ICU constraint.

In the epidemiology literature there exists a series of papers with
analytical solutions for optimal policy.\footnote{%
There is a much larger literature studying epidemics without controls, of
course, see for instance Dickison et al (2012) and Brauer and
Castillo-Chavez (2011), Pastor-Satorras et al. (2015) and references therein.%
} For a literature review on the early research see Wickwire (1977). Many
papers model vaccinations (Morton and Wickwire, 1974; Ledzewicz and Sch\"{a}%
ttler, 2011; Hu and Zou, 2014, Laguzet and Turinici, 2015; Maurer and de
Pinho, 2015), some model screening (Ainseba \& Iannelli, 2012). The previous
papers focusing on suppression (or quarantine) either restrict the policy
(e.g., diLauro et al., 2020, see also Nowzari 2016 for a review) or abstract
from the fact that increasing the suppression is costly, obviously a key
aspect of any economic analysis.\footnote{%
Many papers analytically solve for a suppression policy while respecting a
budget constraint (Hansen and Day, 2011; Bolzoni et al., 2019) or a time
constraint (so that the suppression cannot be too long, Morris et al 2020)
but disregarding that more suppression within a time period is costlier than
less suppression (Bolzoni et al., 2017; Piunovskiy et al., 2019). This is
isomorphic to restricting the suppression policy to be binary since, once
there is suppression within a time period, it may as well be at full force.
We allow the suppression policy to take any value within a period and change
in any way between time periods. Grigorieva et al. (2016) and Grigorieva and
Khailov (2014) analyze an objective of minimizing the number of infectious
during or at some end period, but the control bears no cost. Abakus (1973)
and Behncke (2000) analyzes an objective of minimizing the total (over time)
number of infected quarantine (see his Section 3) but the cost of putting a
person in quarantine is only taken once so is independent of the length of
quarantine. Finally, Gonzales-Eiras and Niepelt (2020) analyze an S.I.
model, finding, just like some of the papers above, that the optimal control
is binary.} The paper closest to ours is an elegant analysis by Kruse and
Strack (2020). They also look at optimal suppression with costs which are
increasing in suppression. They show existence of an optimizer for a rather
general health-cost function but only solve for the optimizer in the special
case where the health costs are linear in the number of currently infected.%
\footnote{%
They also show existence of an optimizer for when a vaccine can arrive.}
This is equivalent to assuming that the total number of deaths (over time)
is proportional to the total number of infected (the linearity assumption
implies bang-bang solutions for suppression) so it does not (directly)
matter how many are infected at the same time like is the focus of our paper
(in that sense, it is similar to Grigorieva et al., 2016, and Grigorieva and
Khailov, 2014). Our contribution is thus complementary to theirs since we
study a health cost which specifically captures the overwhelming of the
health-care system.

\section{Model}

Our model setup closely follows the canonical Susceptible-Infectious-Removed
model (Kermack and McKendrick 1927; see also Brauer and Castillo-Chavez,
2011, for an excellent overview). At any time $t\geq 0$, let $x\left(
t\right) $ be the population share of individuals who at time $t$ are
susceptible to the infection, and let $y\left( t\right) $ be the population
share of individuals who are infected at time $t$. All infected individuals
are assumed to be contagious, and population shares are defined with respect
to the initial population size,\textbf{\ }$N$\textbf{. }Let $\lambda \left(
t\right) $ be the rate at time $t$ of pairwise meetings between susceptible
and infected, and let $q\left( t\right) $ be the probability of contagion
when an infected person meets a susceptible person at time $t$. Write $%
b\left( t\right) =\lambda \left( t\right) q\left( t\right) $. Infected
individuals are removed from the population at rate $\alpha >0$.\footnote{%
Implicitly this assumes that the duration of the infection in an individual
is an i.i.d. exponentially distributed random variable with mean value $%
1/\alpha $.} This may be either because they get immune or because they die.
An important assumption is that those who are infected never again become
susceptible.

The population dynamic is then defined by the following simple system of
ordinary differential equations: 
\begin{equation}
\left \{ 
\begin{array}{l}
\dot{x}\left( t\right) =-b\left( t\right) y\left( t\right) x\left( t\right)
\\ 
\dot{y}\left( t\right) =b\left( t\right) y\left( t\right) x\left( t\right)
-\alpha y\left( t\right)%
\end{array}%
\right.  \label{ODE3}
\end{equation}%
The initial condition is $x\left( 0\right) =1-\varepsilon $ and $y\left(
0\right) =\varepsilon $, for some $\varepsilon \in \left( 0,1\right) $. That
is, the infection enters the population at time zero in a population share $%
\varepsilon >0$. The state space of this dynamic is $\Delta =\left \{ \left(
x,y\right) \in \mathbb{R}_{+}^{2}:x+y\leq 1\right \} $. The only difference
from the standard S.I.R. model is that the propagation coefficient $b\left(
t\right) $, instead of being a constant over time, say, $b\left( t\right)
=\beta >0$ for all $t\geq 0$, we here allow it to vary over time.\footnote{%
If $z\left( t\right) $ denotes the population share of removed individuals
in a standard S.I.R. model, then its dynamic is $\dot{z}\left( t\right)
=\alpha y\left( t\right) $, and $x\left( t\right) +y\left( t\right) +z\left(
t\right) =1$ at all times $t\geq 0$.}

Indeed, we will view $b:\mathbb{R}_{+}\rightarrow \mathbb{R}_{+}$ as a
function in the hands of a social planner who strives to minimize the
economic and social costs of shutting down parts of the economy and social
life in the population, while never letting the population share of infected
individuals, $y\left( t\right) $, exceed an exogenously given level $\gamma $%
. The latter is interpreted as the capacity of the health-care system to
treat infected patients. We refer to it as the \textit{ICU capacity} or 
\textit{constraint}.\footnote{%
If, for example, on average 20\% of those infected need intense care and the
number of ICUs is $C$ in a population of size $N$, then $\gamma =5C/N$. To
allow for risk, the ICU constraint can of course also include a margin to
the actual limit.} It is meant to capture a situation, such as under
Covid-19, where if the number of simultaneously and seriously ill exceeds
the number of respirators implies instant death. Not breaching the ICU
capacity thus ensures that all get the best possible care.\footnote{%
In practice this of course is no guarantee against fatalities. We implicitly
assume that those that pass away despite getting the best care are not
within the control of the policy maker. See, e.g., Kruse and Strack (2020),
Grigorieva et al. (2016) and Grigorieva and Khailov (2014) for models where
the objective is to minimize the number of infected.}

We assume that the cost of keeping $b\left( t\right) $ below its \textit{%
natural}, or \textit{unregulated level} $\beta $ is a linear function of the
difference, while there is no cost of moving $b\left( t\right) $ above $%
\beta $. The latter assumption is made to "tilt the table" against us in the
subsequent analysis, where we will show that even under this assumption, it
is suboptimal to enhance the propagation of the infection even if this can
be done at no cost. Formally, the cost function $\mathcal{C}:\mathcal{B}%
\rightarrow \mathbb{R}_{+}$ is defined by $\mathcal{C}\left( b\right)
=\int_{0}^{\infty }\left[ \beta -b\left( t\right) \right] _{+}dt$, and the
social planner faces the optimization program%
\begin{equation}
\min_{b\in \mathcal{B}_{\gamma }}\mathcal{C}\left( b\right) ,  \label{OPT3}
\end{equation}%
where $\mathcal{B}$ is the class of piecewise continuous functions $b:%
\mathbb{R}_{+}\rightarrow \mathbb{R}_{+}$ that have finitely many points of
discontinuity (including no discontinuity at all), and $\mathcal{B}_{\gamma
} $, for any given $\gamma >0$, is the subset of functions in $\mathcal{B}$
for which $y\left( t\right) \leq \gamma $ at all times $t\geq 0$.\footnote{%
To be more precise, we require that there is a finite set $T\subset \mathbb{R%
}_{+}$ such that the function $b:\mathbb{R}_{+}\rightarrow \mathbb{R}_{+}$
is continuous at all other points, and that it is everywhere left-continuous
and has a right limit. We also require that $b$ is positive everywhere,
except on at most finitely many connected components.}

We focus on situations in which $\varepsilon <\gamma $, that is, when the
initial infection level is below the ICU capacity constraint. Moreover, we
assume that $\beta >\alpha $. Otherwise the population share of infected
individuals does not increase from its initial value, which would imply herd
immunity already from the outset, and thus the social planner's optimization
program then has a trivial solution; \textit{laissez-faire}, that is, $%
b\left( t\right) \equiv \beta $.

\section{Analysis}

We first comment on the set of policies allowed by the optimization program.
For functions $b\in \mathcal{B}$, it can be shown that (\ref{ODE3}) defines
a unique solution trajectory through any given state $\left( x\left(
t_{0}\right) ,y\left( t_{0}\right) \right) \in \Delta $ and time $t_{0}\geq
0 $.\footnote{%
See Appendix for a proof. For this class of functions $b$, the time
derivatives in (\ref{ODE3}) represent left derivatives. The solution
trajectories $\left( x\left( t\right) ,y\left( t\right) \right) _{t>0}$ are
then uniquely determined and are continuous in $t$.} Trivially all constant
functions $b:\mathbb{R}_{+}\rightarrow \mathbb{R}_{+}$, with $b\left(
t\right) =\delta $ for some $\delta >0$, belong to $\mathcal{B}$. However,
they do not all belong to $\mathcal{B}\left( \gamma \right) $, i.e., they
may violate the ICU constraint. It is easy to show that such constant
policies belong to $\mathcal{B}\left( \gamma \right) $ if $\delta $ is
sufficiently low, for any given $\gamma >0$. Thus, to choose $\delta $ as
high as possible, while keeping $y\left( t\right) \leq \gamma $ for all $%
t\geq 0$, is a feasible policy (belongs to $\mathcal{B}\left( \gamma \right) 
$), and can be called \textit{flattening the curve}. However, such a policy
incurs an infinite cost since it lasts forever, if and only if $\delta
<\beta $. An alternative feasible control function, with finite cost, is to
only temporarily keep $b\left( t\right) $ at a constant level $\delta <\beta 
$ over a carefully chosen finite time interval, where $\delta $ is such that 
$y\left( t\right) \leq \gamma $ for all $t\geq 0$. However, as will be shown
below, also such \textquotedblleft temporary constant shut
down\textquotedblright \ policies are suboptimal. Before turning to the
formal statement of our main result, we analyze some general properties of
the dynamic induced by (\ref{ODE3}).

\subsection{The dynamic}

Some well-known properties of the solutions to standard S.I.R. models hold
also here (see Brauer and Castillo-Chavez, 2011). A key such property is
that the population share of susceptible individuals, $x\left( t\right) $,
is non-increasing over time $t$. Roughly speaking, this follows from the
first equation in (\ref{ODE3}), since $b\left( t\right) $ is always
non-negative and $y\left( t\right) $ is positive at all times $t\geq 0$.
Being bounded from below by zero, $x\left( t\right) $ necessarily has a
limit value as $t\rightarrow \infty $, which we denote $x_{\infty }$.
According to (\ref{ODE3}), also the sum $y\left( t\right) +x\left( t\right) $
is strictly decreasing over time $t$, and hence also this sum has a limit
value, $x_{\infty }+y_{\infty }$. By standard arguments, it is easily
verified that this implies that $y_{\infty }=0$.\footnote{%
If $y_{\infty }>0$, then $x\left( t\right) +y\left( t\right) \rightarrow
-\infty $.} In other words, in the very long run, the population share of
infected individuals tends to zero. Denoting by $z_{\infty
}=\lim_{t\rightarrow \infty }z\left( t\right) $ the total population share
of removed individuals during the whole epidemic, we thus have $z_{\infty
}=1-x_{\infty }$, and $Nz_{\infty }$ is approximately (for large $N$), the
total number of infected individuals during the epidemic.

Let us now consider the solution to (\ref{ODE3}) through any through any
given state $\left( x\left( t_{0}\right) ,y\left( t_{0}\right) \right) \in
\Delta $ and time $t_{0}\geq 0$, where $0<x\left( t_{0}\right) <1$ and $%
0<y\left( t_{0}\right) <1$. Dividing both sides of the second equation in (%
\ref{ODE3}) by $x\left( t\right) >0$ and integrating, one obtains%
\begin{equation*}
\ln x\left( t\right) =x\left( t_{0}\right) -\int_{t_{0}}^{t}b\left( s\right)
y\left( s\right) ds\quad \forall t\geq t_{0}\text{.}
\end{equation*}%
Moreover, integrating the sum of the two equations in (\ref{ODE3}), we
obtain 
\begin{equation*}
x\left( t\right) +y\left( t\right) =x\left( t_{0}\right) +y\left(
t_{0}\right) -\alpha \int_{t_{0}}^{t}y\left( s\right) ds\quad \forall t\geq
t_{0}\text{.}
\end{equation*}

\subsubsection{Constant policy}

In particular, if $b\left( t\right) =\delta >0$ for all $t\geq t_{0}$, for
some $\delta >0$, then for all $t\geq t_{0}$: 
\begin{equation*}
\ln \frac{x\left( t\right) }{x\left( t_{0}\right) }=-\delta
\int_{t_{0}}^{t}y\left( s\right) ds=\frac{\delta }{\alpha }\left[ x\left(
t\right) -x\left( t_{0}\right) +y\left( t\right) -y\left( t_{0}\right) %
\right] ,
\end{equation*}%
or 
\begin{equation}
y\left( t\right) =y\left( t_{0}\right) +\frac{\alpha }{\delta }\ln \left( 
\frac{x\left( t\right) }{x\left( t_{0}\right) }\right) -x\left( t\right)
+x\left( t_{0}\right) \quad \forall t\geq t_{0}\text{.}  \label{orb}
\end{equation}%
This equation is well-known for S.I.R. models. Moreover, (\ref{orb}) implies
that $\left( x\left( t\right) ,y\left( t\right) \right) \rightarrow \left(
0,x_{\infty }\right) \in \Delta $, where $x_{\infty }$ by continuity solves (%
\ref{orb}) for $y\left( t\right) =0$, so%
\begin{equation}
x_{\infty }=\frac{\alpha }{\delta }\ln \frac{x_{\infty }}{x\left(
t_{0}\right) }+x\left( t_{0}\right) +y\left( t_{0}\right) .  \label{xoo}
\end{equation}%
Since $x\left( t\right) $\ is strictly decreasing, $x_{\infty }<x\left(
t_{0}\right) $. It is easily verified that the fixed-point equation (\ref%
{xoo}) has a unique solution $x_{\infty }\in \left( 0,x\left( t_{0}\right)
\right) $.\footnote{%
The right-hand sides of () is a continuous and strictly increasing functions 
$f:\left( 0,x\left( t_{0}\right) \right) \rightarrow \mathbb{R}$ of $%
x_{\infty }$. Moreover, $f\left( x_{\infty }\right) \rightarrow -\infty $ as 
$x_{\infty }\downarrow 0$ and $f\left( x\left( t_{0}\right) \right) =x\left(
t_{0}\right) +y\left( t_{0}\right) >x\left( t_{0}\right) $, $f^{\prime }>0$
and $f^{\prime \prime }<0$, so there exists a unique fixed point in $\left(
0,x\left( t_{0}\right) \right) $.}

The maximal population share of infected individuals, 
\begin{equation*}
\hat{y}=\sup_{t\geq 0}y\left( t\right)
\end{equation*}%
(still for $b\left( t\right) \equiv \delta \ $for some $\delta >0$) is the
peak infection level. It obtains when $\dot{y}\left( t\right) =0$, or,
equivalently (by (\ref{ODE3})), when $x\left( t\right) =\alpha /\delta $.
From (\ref{orb}) we obtain%
\begin{equation}
\hat{y}=1+\frac{\alpha }{\delta }\ln \left( \frac{\alpha }{\delta \left(
1-\varepsilon \right) }\right) -\frac{\alpha }{\delta }\text{.}  \label{peak}
\end{equation}%
This is thus the maximal population share of infected individuals when the
suppression policy is held constant over time. Once the population share $%
x\left( t\right) $ of susceptible individuals has fallen below the level $%
\alpha /\delta $, achieved precisely when $y\left( t\right) =\hat{y}$, flock
immunity is obtained; the population share $y\left( t\right) $ of infected
individuals falls. In particular, the limit state as $t\rightarrow \infty $
is Lyapunov stable. That is, there is no risk of a second infection wave,
since after any small perturbation of the limit population state $\left(
x_{\infty },0\right) \in \Delta $, obtained by exogenously inserting a small
population share of infected individuals, the population share of infected
individuals will fall gradually back towards zero, while the population
share of susceptible individuals gradually moves towards a somewhat lower,
new limit value.

Equation (\ref{peak}) is particularly relevant for the case when $\delta
=\beta $, that is, under under laissez-faire. Because if the peak of the
infection wave then does not exceed the ICU capacity constraint, that is, if 
\begin{equation}
1+\frac{\alpha }{\beta }\ln \left( \frac{\alpha }{\beta \left( 1-\varepsilon
\right) }\right) -\frac{\alpha }{\beta }\leq \gamma ,  \label{LF}
\end{equation}%
then laissez-faire is optimal; $b^{\ast }\left( t\right) \equiv \beta $
solves (\ref{OPT3}) at no cost. But if the peak is above the ICU constraint,
regulation has to be implemented. This is the topic of the next subsection.

\subsection{Optimization}

To the best of our knowledge, the optimization program (\ref{OPT3}) has not
been analyzed before. We summarize below our main result, which treats all
cases when laissez-faire is suboptimal. If (\ref{LF}) does not hold, then
the solution orbit (\ref{orb}) under laissez-faire intersects the capacity
constraint $y\left( t\right) =\gamma $ twice. Let $\tau _{1}>0$ be the first
such time and let $x\left( \tau _{1}\right) $ be the population share of
susceptible individuals at that time. Then 
\begin{equation*}
\tau _{1}=\min \left \{ t\geq 0:x\left( t\right) =1-\gamma +\frac{\alpha }{%
\beta }\ln \left( \frac{x\left( t\right) }{1-\varepsilon }\right) \right \}
\end{equation*}%
(where $x\left( t\right) $ is solved for according to (\ref{ODE3}) when $%
b\left( t\right) \equiv \beta $), and $x\left( \tau _{1}\right) $ is the
larger of the two solutions to the associated fixed-point equation in $x$,%
\begin{equation}
x=1-\gamma +\frac{\alpha }{\beta }\ln \left( \frac{x}{1-\varepsilon }\right)
.  \label{x}
\end{equation}%
We note that $x\left( \tau _{1}\right) >\alpha /\beta $.\footnote{%
This follows from the observation that the derivative of the right-hand side
of (\ref{x}) is less than unity at $x=x\left( \tau _{1}\right) $.} Let 
\begin{equation*}
\tau _{2}=\tau _{1}+\frac{1}{\alpha \gamma }\left( x\left( \tau _{1}\right) -%
\frac{\alpha }{\beta }\right)
\end{equation*}

\begin{theorem}
\label{Thm: Main}Suppose that $\varepsilon <\gamma $, $\alpha <\beta $ and (%
\ref{LF}) does not hold. There exists a solution to program (\ref{OPT3}),
one of which is the policy $b^{\ast }\in \mathcal{B}\left( \gamma \right) $
defined by%
\begin{equation*}
b^{\ast }\left( t\right) =\left \{ 
\begin{array}{ll}
\beta & \text{for }t\leq \tau _{1} \\ 
\frac{\beta }{1+\alpha \beta \gamma \left( \tau _{2}-t\right) } & \text{for }%
\tau _{1}<t\leq \tau _{2} \\ 
\beta & \text{for }t>\tau _{2}%
\end{array}%
\right.
\end{equation*}%
Every optimal policy $b\in B\left( \gamma \right) $ agrees with $b^{\ast }$
on $\left[ 0,\tau _{2}\right] $ and satisfies $b\left( t\right) \geq \beta $
for all $t>\tau _{2}$.
\end{theorem}

\bigskip

We note that the optimal policy is laissez-faire both before time $\tau _{1}$
and after time $\tau _{2}$. We also note that the optimal policy has exactly
one discontinuity, namely, a sudden shut-down at time $\tau _{1}$; then $%
b^{\ast }\left( t\right) $ falls from $b^{\ast }\left( \tau _{1}\right)
=\beta $ to 
\begin{equation*}
\lim_{t\downarrow \tau _{1}}b^{\ast }\left( t\right) =\frac{\beta }{1+\alpha
\beta \gamma \left( \tau _{2}-\tau _{1}\right) }=\f{\alpha}{x(\tau_1)}.
\end{equation*}

From time $\tau _{1}$ on, $b^{\ast }\left( t\right) $ rises continuously
until time $\tau _{2}$, at which point $b^{\ast }\left( t\right) $ reaches
the level $\beta $. In the mean-time, between times $\tau _{1}$ and $\tau
_{2}$, the population share $y\left( t\right) $ of infected individual
remains constant, at the capacity level $\gamma $, while the population
share $x\left( t\right) $ falls linearly over time to the level $\alpha
/\beta $, reached at time $\tau _{2}$.

One obtains the following expression for the minimized cost:%
\begin{eqnarray} \label{I}
\nonumber\mathcal{C}\left( b^{\ast }\right) &=&\beta \int_{\tau _{1}}^{\tau _{2}}%
\frac{\alpha \beta \gamma \left( \tau _{2}-t\right) }{1+\alpha \beta \gamma
\left( \tau _{2}-t\right) }dt\\
\nonumber&=&\frac{1}{\alpha\gamma }\int_{0}^{\alpha \beta \gamma \left( \tau _{2}-\tau
_{1}\right) }\left( 1-\frac{1}{1+s}\right) ds \\
 \nonumber&=&\frac{1}{\alpha\gamma }\left[ \alpha \beta \gamma \left( \tau _{2}-\tau
_{1}\right) -\ln \left( 1+\alpha \beta \gamma \left( \tau _{2}-\tau
_{1}\right) \right) \right] 
\\
 &=&\f1\gamma\lt(x(\tau_1)-\f\alpha\beta\ln(x(\tau_1))\rt)+\f\alpha{\beta\gamma}\lt(\ln\lt(\f{\alpha}\beta\rt)-1\rt)
\end{eqnarray}

The proof of Theorem \ref{Thm: Main} is mathematically involved, and is
given in the Appendix. It uses measure theory and views the minimization as
taking place in phase space (much in line with equation (\ref{orb})). For a
rich enough measure space, existence of a solution to (\ref{OPT3}) is
obtained by topological arguments. Invoking the Picard-Lindel\"{o}f theorem,
it is shown that the differential equations (\ref{ODE3}) indeed uniquely
define solutions. The next step in the proof is to show that the minimizer
measure is absolutely continuous (with respect to Lebesgue measure). This
brings us back to functions $b\in \mathcal{B}$, now viewed as transforms of
Radon-Nikodyn derivatives of the measures in question. The rest of the proof
consists in verifying that the above function, $b^{\ast }$, indeed
corresponds to an optimal measure, and that it is unique in the sense
stated. In particular, one needs to show that it is neither worthwhile to
slow down nor speed up the infection in its early phase (before time $\tau
_{1}$). An early slow-down would postpone the problem at a cost but without
benefit, and an early speed-up, although costless in our model, would imply
that the capacity constraint is reached sooner and at a higher speed, which
would require an even more drastic, and costly, shut down when the capacity
constraint is reached.

\begin{figure}[h]
  \includegraphics[width=12cm]{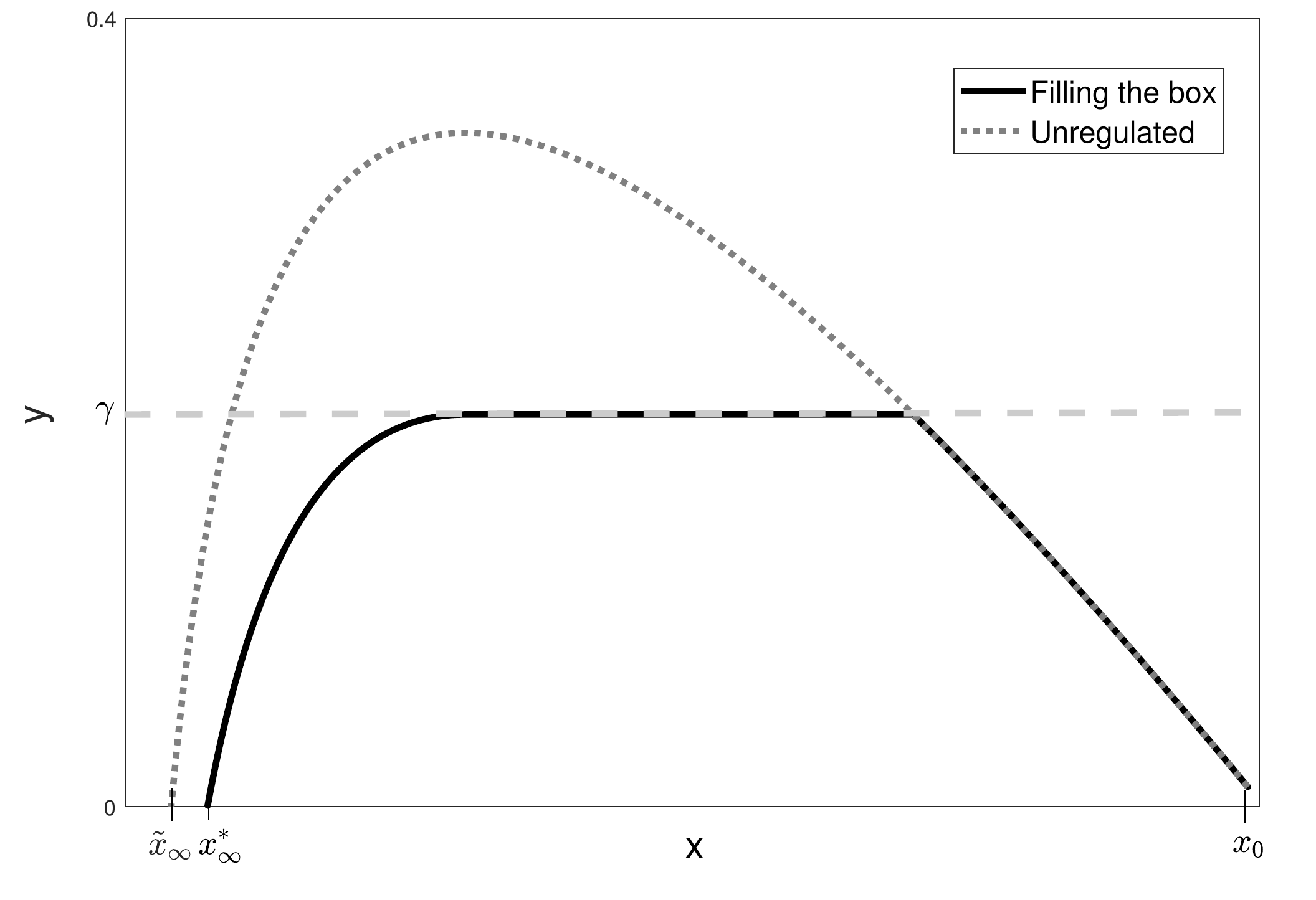}
   \caption{The solution orbit (solid) in the $(x,y)$-plane under the optimal policy $b^*$, and the solution orbit under unregulated spread (dotted)
   . Parameter values used: $\protect \alpha =0.3$, $\protect \beta %
=1$, $\protect \gamma =0.2$, and $\protect \varepsilon =0.01$.}
\label{Fig: Orbit optimal}
  \end{figure}

The result is illustrated in Figure \ref{Fig: Orbit optimal}, where the
solid kinked curve is the solution orbit induced by (\ref{ODE3}) under the
optimal control function $b^{\ast }$. The dotted curve is the infection
orbit under laissez-faire ($b\left( t\right) \equiv \beta $). We note that
the limit share of susceptible individuals, $x_{\infty }$, is higher under
the optimal policy than under laissez-faire. Recalling that $z_{\infty
}=1-x_{\infty }$ is the total population share of infected individuals
during the epidemic, we conclude that, the policy not only respects the ICU
constraint, it also indirectly affects the total number that will ultimately
have been infected at some point in time.

\begin{figure}[h]
  \includegraphics[width=12cm]{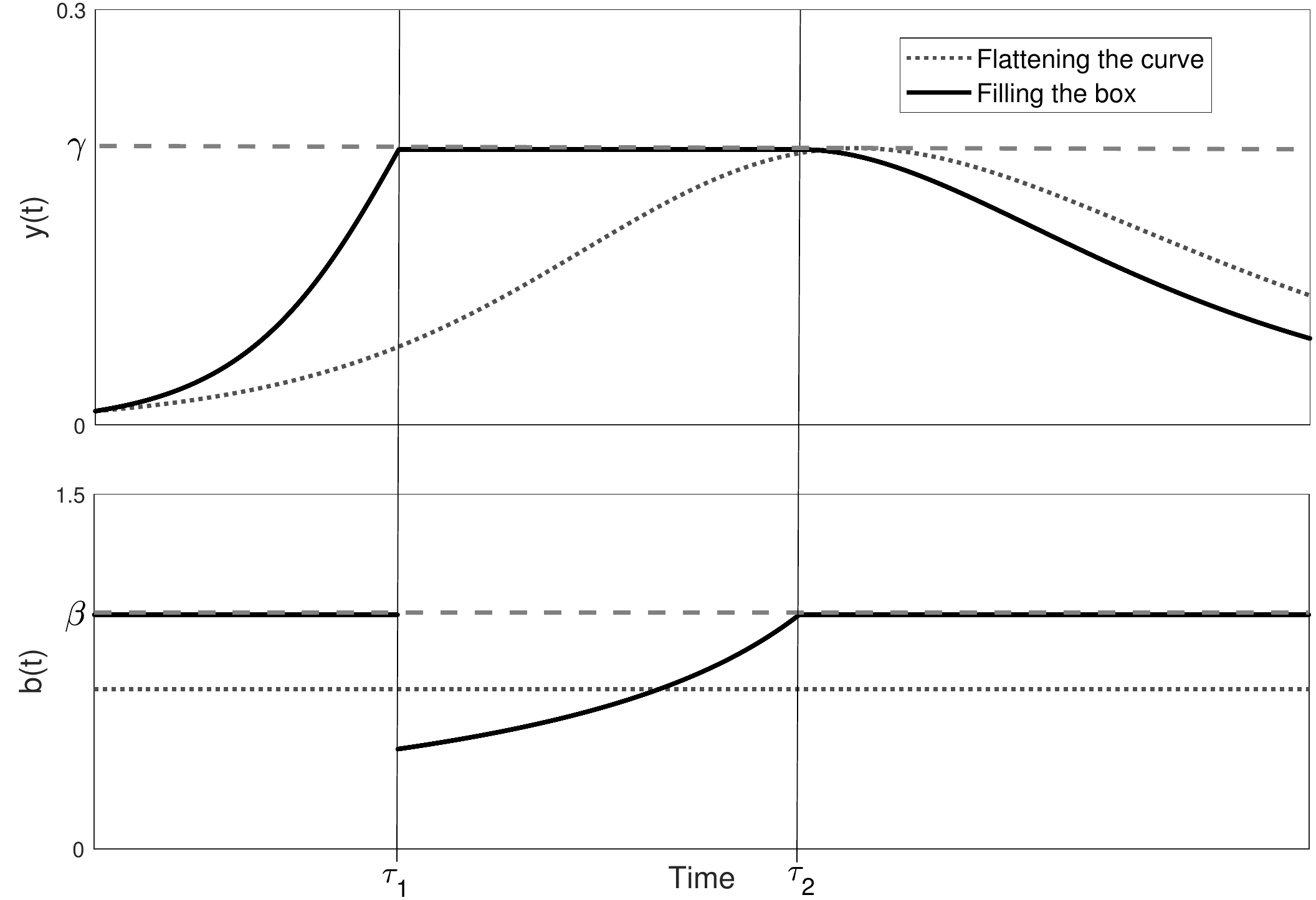}
   \caption{Upper panel: The share of infected
over time under the optimal policy (solid) and flattening the curve
(dotted). The horizontal dashed line represents the ICU constraint $\protect%
\gamma $. Lower panel: Optimal suppression (solid) and flattening-the-curve
suppression (dotted). The horizontal dashed represents the baseline spread $%
\protect \beta $. Parameter values used: $\protect \alpha =0.3$, $%
\protect \beta =1$, $\protect \gamma =0.2$, and $\protect \varepsilon =0.01$.}
\label{Fig: Time dynamics}
  \end{figure}

Figure \ref{Fig: Time dynamics} depicts the optimal policy as a function of
time in comparison to a strategy of flattening the curve, here assumed to
take the form: keep $b\left( t\right) $\ at at the level $\delta <\beta $\
for which $\hat{y}=\gamma $ (see (\ref{peak})) until the infection wave has
reached its peak, and then return to laissez-faire, $b\left( t\right) =\beta 
$ (outside the time range of the figure). The upper panel shows the dynamics
of infections and the lower panel the policy $b\left( t\right) $ . 

As can be seen, and as
expressed by the theorem, the optimal policy is characterized by leaving the
spread unregulated initially, then a sudden shut-down of society (a
discontinuity at $\tau _{1}$), followed by gradual (continuous) opening of
society, until $\tau _{2}$, from which onwards the propagation is not
regulated. The time axis and the ICU constraint create a square -- a box.
The economic logic behind the optimal policy is essentially to ensure that
we do not close down society while leaving idle ICU resources --
\textquotedblleft filling the box\textquotedblright . This implies that
whenever the natural spread is not threatening the constraint, it should go
unregulated. This holds in the early phase when only few have been infected,
and in the last phase, when many have already been infected but most of them
also have recovered. It is only when the epidemic may breach the ICU
constraint -- the second phase -- that it should be regulated. In order to
ensure that the constraint is not breached, strong suppression has to be
imposed when reaching the ICU constraint -- a sudden shutdown. The reason
for the abruptness of this policy (the discontinuity) is that the natural
infection is progressing very quickly at that point, so a sudden break is
needed to stop it. This can be seen in the lower panel by the drop at $\tau
_{1}$. After that, $b^{\ast }$ gradually increases. The reason for this is
that the suppression only needs to keep the infection just below the ICU
constraint. Then since over time the number of susceptible ($x$) is falling,
the number of infected ($y$) is held endogenously constant and since new
infections depends on their product ($b\left( t\right) y\left( t\right)
x\left( t\right) $) it follows that $b^{\ast }$ is increasing during the
second phase. The policy as a function of the population share simply is $%
b^{\ast }\left( t\right) =\alpha /x\left( t\right) $, i.e., recovery ($%
\alpha $) determines what share of the susceptible population that can be
allowed to be infected. A few further remarks about optimal policies are now
in place.

It may be noted that the optimal policy never attempts to fully eradicate
the spread. In our model, like in all standard S.I.R. models, this is since $%
y$ only asymptotically goes to zero. Hence, full eradication (a form of
extreme corner solution) would imply locking down forever. We discuss this
further in the conclusions.

Furthermore, the optimal policy is unique during the first and the second
phase but not during the third. The uniqueness during the first phase is not
obvious.\ To see this note that here there is no reason to hold back the
spread. Then, given that $b>\beta _{0}$ has been assumed to be costless, why
would accelerating the spread not be optimal? The answer is that, if one
does that, then the ICU capacity is reached at a high speed of infection
hence it would require hitting the breaks very hard. This is not optimal.
The multiplicity of optimal strategies during the third is due to the same
assumption -- acceleration is free. Hence, not\ only laissez-faire is
optimal, but also acceleration (of which we can think of as stimulus for
economic interaction). The acceleration cannot be too fast, however, as it
may then breach the ICU constraint. Naturally, should we assume that there
is a cost of acceleration (even the slightest) this multiplicity disappears
and a unique optimal policy emerges also in the third phase -- laissez-faire.

Compared with the optimal policy, \textquotedblleft flattening the
curve\textquotedblright \ implies costs that lead to idle resources. This is
visible in the upper panel of Figure \ref{Fig: Time dynamics}, where costs
are incurred without the spread posing a threat to the health system -- both
before and after the peak, suppression costs are incurred for no reason. The
additional cost of flattening the curve (instead of filling the box) can be
seen in the lower panel by comparing the rectangle between $\beta $ and the
dashed-dotted line on the one hand with the area between $\beta $ and the
solid line on the other. It is potentially very large, in particular if the
policy maker continues to flatten the curve long after the peak.

\section{Concluding discussion}

This paper has developed an economic SIR model to provide an \textit{%
analytical} answer to the question: What is the optimal time-varying
suppression policy to avoid a collapsed health-care system when suppression
is costly? We have shown that the general recommendation of
\textquotedblleft flattening the curve\textquotedblright \ is suboptimal.
Instead the optimal policy essentially prescribes \textquotedblleft filling
the box\textquotedblright : in an initial phase the spread is unregulated
until the number of infected approaches the ICU constraint whereby, in a
second phase, suppression discontinuously increases and then gradually drops
until, in a third phase, the spread is left unregulated again.

A contribution of the paper is methodological, showing how to obtain a fully
analytical solution to an S.I.R. model with economic costs which are
increasing in suppression. Another contribution is the policy implication of
\textquotedblleft filling the box\textquotedblright . We discuss here the
robustness of this policy to various perturbations.

In our model attempting for a complete wipe out of the spread is never
optimal. Technically this is since the number of infected only
asymptotically goes to zero hence a wipe out would require suppression to be
in place for the infinite future. Naturally, if the rate of recovery from
infection happens quickly for everyone it could be optimal to go for a full
wipe out right away. We do not, however, find that feasible in most cases
since it is hard to practically identify all infected and since in practice
the cost of full suppression ($b=0$) is virtually infinite -- after all,
people need to access food and medical services. Furthermore, unless
countries are closed more or less indefinitely, a very costly prospect,
under a pandemic such as Covid-19 one would be bound to import new cases.

In the model we have assumed that the only medical harm is if violating the
ICU constraint. If two assumptions were added -- \ medical harm from the
aggregate number of infected and existence of a vaccine within a reasonable
time frame -- then suppressing the spread more than what our policy
prescribes could be optimal. Likewise if the number of simultaneously
infected would cause harm more gradually.\ This, however, does not seem to
be case for Covid-19 where the bottleneck in most countries is the number of
respirators -- \ respecting the ICU constraint is the main issue. Another
factor that \textit{could} suggest early suppression is if the ICU
constraint can be expanded (for Covid-19 equivalent to an increased number
of respirators or development of a cure or improved treatment). Another
possibility is that one learns about the parameters. However, for that to
motivate regulation early, one has to assume that the suppression itself
does not distort the signal.\ Finally, if the cost of suppression were
convex, in particular so that small suppression is very cheap, then that
would motivate some suppression early on. However, it would still most
probably be optimal to discontinuously increase suppression to ensure that
the number of infected is constant just below the ICU constraint.

This discussion highlights that there exist a number of questions that call
for an analytical approach for full understanding of their impact. Our model
and tools of analysis provide a stepping stone for doing so.


\pagebreak

\section{Appendix: Proof of Theorem \ref{Thm: Main}}

This appendix provides a proof of Theorem \ref{Thm: Main}, namely that $b^*$ is a global minimizer for the control
problem of optimizing the functional $\cC$ given just above \eqref{OPT3}, and that any other global minimizer
coincides with $b^*$ up to time when $x$ hits $\alpha/\beta$.\par\me
Our strategy consists of the following steps: 
\begin{itemize}
\item The optimization problem is written in the phase space $\tr$.
\item The new formulation admits a natural extension on 
 a signed-mea\-sure space.
 \item Topological properties of this measure space and of the functional deduced from $\cC$
imply the existence of a global minimizer.
\item 
A priori a global minimizer is a general signed measure, but it turns out to be 
an absolutely continuous, bringing us back to a  functional setting.
\item 
Calculus of variation arguments show that the minimizer is uniquely determined until the time when $x$ reaches the level $\alpha/\beta$, and this leads to  Theorem \ref{Thm: Main}.
\end{itemize}
\par
To the best of our knowledge, there is no such result in  calculus of variations or optimal control theory (see the books of Clarke \cite{MR3026831} or
 Liberzon \cite{MR2895149}).
We therefore give a direct and self-contained proof (only requiring a first knowledge of measure theory, as  can be found e.g.\ in Rudin \cite{MR924157}). For the optimization problem at hand, our extension to measure spaces seems natural, and we believe it is original. The Euler-Lagrange equations will not be satisfied  and the constraints will play a more important role.
This is related to the fact that we consider a cost of suppression that is linear in downwards deviations, and zero for upwards deviations. 
If one is interested in more general costs of the form
\bqn{wiC}
\wi \cC(b)&\df&\int_0^{+\iy} F(\beta- b(t))\,dt\eqn
where the mapping $F\st\RR\ri\RR_+$ is e.g.\ a strictly convex function attaining its minimum at 0, then the Euler-Lagrange equations admit solutions leading
to optimal policies different from $b^*$ (but $b^*$ remains a minimizer for certain functionals $F$, see Remark \ref{r2} at the end of this appendix). When $F$ is close to the mapping $(\cdot)_+$ considered here, 
for
instance if $F$ is given by
\bq
\fo x\in\RR,\qquad F(x)&\df&\lt\{\begin{array}{ll}
x^{1+\epsilon}&\hbox{, when $x\in\RR_+$}\\
\epsilon \vert x\vert^{1+\epsilon}&\hbox{, when $x\in\RR_-$}\end{array}\rt.\eq
where $\epsilon >0$ is small,
we expect that the corresponding
solution will be close to $b*$. In particular a jump will still occur.
We plan to investigate more precisely this situation in  future work. \par
Let us now move toward the proof of Theorem \ref{Thm: Main} according to the above strategy.
We assume in the sequel that we are in the ``interesting'' case where $y_0<\gamma$ and where the \textit{laissez-faire} policy $b\equiv\beta$
leads $y$ to take values strictly large than $\gamma$ ($y_0=\varepsilon$ in the main text). This hypothesis will be referred to as the \textit{underlying assumption}.

\subsection{The  phase space $\tr$}\label{txps}

We begin by rewriting the constrained control problem of minimizing $\cC$ on $\cB_{\gamma}$ as a optimization problem in the associated phase space $\tr$.\par\sm
Let us be more precise:  $\cB$ is the set of piecewise continuous functions $b\st\RR_+\ri\RR_+$ with a finite number of discontinuities and such that $\{t\geq 0\st b(t)=0\}$ has a finite number of connected components.
Let $(x_0,y_0)\in\tr$, with $y_0\in (0,1)$, as well as $b\df(b(t))_{t\geq 0}\in\cB$ be given and consider $(x,y)\df (x(t),y(t))_{t\geq 0}$ the maximal (over time) solution of the S.I.R.\  ODE\ \eqref{ODE3}
starting from $(x(0),y(0))=(x_0,y_0)$ (this is a slight generalization of the setting of Theorem \ref{Thm: Main}, where $x_0=1-y_0$, the important hypothesis is that the underlying assumption holds).
The existence and uniqueness of this solution is a consequence of the Picard-Lindel\"of or Cauchy-Lipschitz theorem,
extended to a time-dependent vector field that is left continuous with right limits (instead of continuous). The important fact being that the r.h.s.\ of \eqref{ODE3} is locally Lipschitz with respect to $(x(t),y(t))$. By $\cB_{\gamma}$, we designate the set of $b\in\cB$ such that $y$ always remains below $\gamma$.
\par
We begin with a simple observation.
\begin{lem}\label{lem1}
When $y_0\in(0,1)$, the solution $(x,y)$ is defined for all times $t\geq 0$ and
 $(x(t),y(t))\in (0,1)^2\cap\tr$.
\end{lem}
\prooff
As already mentioned, the solution $(x,y)$ is locally unique. From this it follows that
 if $x$ reaches $0$, then it will stay there forever afterward, 
since the r.h.s.\ of  \eqref{ODE3} is zero if $x(t)=0$. Similarly, if $y$ reaches $0$, then it will stay there forever afterward.
It follows that $x$ and $y$ will stay non-negative. As a consequence, $x$ is non-increasing and thus will stay below $x_0$ and never hit 1.
From the identity
\bq
\dot{x}(t)+\dot{y}(t)&=&-\alpha y(t)\ \leq\ 0\eq
we deduce that $x+y$ will stay below $y_0+x_0\leq 1$ and in particular $y$ will stay below 1.
The first equation of \eqref{ODE3} then implies that
\bq
\do{x}(t)&\geq & -b(t) x(t)\eq
and Gronwall lemma shows
\bq
x(t)&\geq x_0\exp\lt(-\int_0^t b(s)\, ds\rt)\eq
so $x$ remains positive. The inequality $x+y\leq 1$ then insures that $y$ never reaches 1.
Finally, the second equation of \eqref{ODE3} then implies that
\bq
\do{y}(t)&\geq & -\alpha y(t)\eq
so that $y(t)\geq y_0\exp(-\alpha t)$ and $y$ cannot reach 0 in finite time.\par
Since $(x,y)$ stays  in the compact square $[0,1]^2$, the solution of \eqref{ODE3}
is defined for all times.
\wwtbp
\par
\begin{rem}\label{eradication}
In particular, since for all $t\geq 0$ we have $y(t)>0$, a part of the population will always remain  infectious, 
whatever the choice of the policy $b$: it is impossible to entirely eliminate the disease. This feature is due to the fact we are considering continuous populations,
it would not be true for approximating finite random populations.
\end{rem}
\par\sm
Introduce $\cB^+$ the set of $b\in \cB$ that are everywhere positive
(nevertheless, if $t$ is a discontinuity point of $b$, we can have $b(t+)=0$).
\par
When $b\in\cB^+$, Lemma \ref{lem1} and the first equation of \eqref{ODE3} imply that $x$ is decreasing, so $x$ admits  a limit $x_\iy\geq 0$ that
it will never reach.
Another consequence is:
\begin{lem}\label{cB+}
Assume that $b\in\cB^+$. There exists a unique function $\varphi\st (x_\iy, x_0]\ri (0,1)$ such that
\bqn{varphi}
\fo t\geq 0,\qquad y(t)&=&\varphi(x(t))\eqn
\par
The function $\varphi$ is piecewise $\cC^1$, its left and right derivatives exist everywhere and differ only at a finite number of points.
Denoting the right derivative by $\varphi'$, we have 
\bq
\fo r\in (x_\iy, x_0],\qquad \varphi'(r)&>&-1\eq
\end{lem}
\prooff
As observed above, for $b\in\cB^+$, $x$ is decreasing from $\RR_+$ to $(x_\iy,x_0]$.
Since $x$ it continuous, it is a homeomorphism between $\RR_+$ to $(x_\iy,x_0]$. Denote by $\tau$ its inverse, so that
\bqn{tau}
\fo u\in (x_\iy, x_0],\qquad x(\tau(u))&=&u\eqn
\par
Let $t\in\RR_+$ be a time where $b$ is continuous.
Let $u\in (x_\iy, x_0]$ be such that $\tau(u)=t$. We can differentiate \eqref{tau} at $u$ to get
that
\bq
\dot{\tau}(u)&=&\f{1}{\dot{x}(\tau(u))}=
-\f{1}{b(\tau(u))uy(\tau(u))}
\eq
\par
Considering discontinuity time $t$ of $b$, we see that the above relation also holds, if $\dot{\tau}(u)$ is seen as a right derivative (recall $\dot{x}$ is a left derivate).
Furthermore, taking into account that $\tau$ is decreasing, we have the existence of the left limit:
\bq
\lim_{v\ri u_-}\dot{\tau}(v)
&=&-\f{1}{b(\tau(u)+)uy(\tau(u))}
\eq
\par
It leads us to define $\varphi$ via
\bq
\fo u\in (x_\iy, x_0],\qquad \varphi(u)&\df& y(\tau(u))\eq
since this is indeed equivalent to \eqref{varphi}.
Its left and right derivatives exist everywhere as a consequence of the differentiability properties of $y$ and $\tau$.
These left and right derivatives  do not coincide only on a finite number of points, those of the form $x(t)$, where $t\in\RR_+$ is a discontinuity time of $b$.\par
Our conventions insure that \eqref{varphi} can be left differentiated everywhere and that
\bq
\fo t\geq 0,\qquad \dot{y}(t)&=&\varphi'(x(t)) \dot{x}(t)\eq
(recall $\dot{y}$ is a left derivate), namely
\bqn{ppp}
\nonumber\fo t\geq 0,\qquad \varphi'(x(t))&=& \f{\dot{y}(t)}{\dot{x}(t)}
=-1+\f{\alpha}{b(t)x(t)}
>-1\eqn
\par
\begin{rem} \label{inv1}
From the knowledge of $\varphi$ it is possible to reconstruct $b$, at least when $\varphi'$ is Lipschitzian.
Indeed, 
\eqref{ODE3} can be written
\bq\lt\{
 \begin{array}{rcl}
 \dot{x}(t)&=& -b(t)x(t)\varphi(x(t))\\[2mm]
  \varphi'(x(t))\dot{x}(t)&=& b(t)x(t)\varphi(x(t))-\alpha \varphi(x(t))
 \end{array}
 \rt.\eq
which implies that
\bq
 \dot{x}(t)&=& - \varphi'(x(t))\dot{x}(t)+\alpha \varphi(x(t))\eq
 i.e.
 \bq
 \dot{x}(t)&=&\alpha\f{ \varphi(x(t))}{1+\varphi'(x(t))}\eq
\par
So when $\varphi'$ is Lipschitzian, we can solve this ODE\ to reconstruct $x\df(x(t))_{t\geq 0}$.
The trajectory $y\df(y(t))_{t\geq 0}$ is then obtained as $(\varphi(x(t)))_{t\geq 0}$
and $b$ via the formula
\bq
\fo t\geq 0,\qquad b(t)&=&-\f{\dot{x}(t)}{x(t)y(t)}\eq
\par
\end{rem}
\par
This inequality can be translated into $\varphi'>-1$ on $(x_\iy, x_0)$.\wwtbp
\par
To any function $\varphi$ as in the previous lemma, associate the quantity
\bq
\cJ(\varphi)&\df& \int_{x_\iy}^{x_0} L(\xi,\varphi(\xi),\varphi'(\xi))\, d\xi\eq
where for any $(\xi, \chi,\chi')\in (x_\iy,x_0]\times (0,1)\times (-1,+\iy)$,
\bq
L(\xi, \chi,\chi')&\df&\f\beta\alpha \lt(\f{1+\chi'}{\chi}-\f{\alpha}{\beta\xi\chi}\rt)_+\eq
\par
The interest of these definitions is to enable us to write the cost  functional $\cC$ in terms of $\varphi$:
\begin{lem}\label{cB+2}
For $b\in\cB^+$ and with the notations of Lemma \ref{cB+}, we have
\bq
\cC(b)&=& \cJ(\varphi)\eq
\end{lem}
\prooff
Equation \eqref{ppp} enables us to recover $b$ in terms of $\varphi$ and $x$:
\bqn{ppp2}
\fo t\geq 0,\qquad b(t)&=&\f{\alpha}{x(t)(1+\varphi'(x(t)))}\eqn
and we deduce that
\bq
\cC(b)&=& \int_0^\iy\lt(\beta-\f{\alpha}{x(t)(1+\varphi'(x(t)))}\rt)_+\, dt\\
&=& -\int_0^\iy\lt(\beta-\f{\alpha}{x(t)(1+\varphi'(x(t)))}\rt)_+\f1{b(t)x(t)y(t)}\, \dot{x}(t)dt\\
&=&\int_{x_\iy}^{x_0}\lt(\beta-\f{\alpha}{u(1+\varphi'(u))}\rt)_+\f1{b(\tau(u))u\varphi(u)}\, du\eq
where we used the change of variable $t=\tau(u)$, the mapping $\tau$ being defined in \eqref{tau}.
\par
Let us remove the term $b(\tau(u))$ in the latter integral.
Replacing $t$ by $\tau(u)$, we get from \eqref{ppp2}
\bq
b(\tau(u))&=&\f{\alpha }{u(1+\varphi'(u))}\eq
so 
that
\bq
\cC(b)&=&\int_{x_\iy}^{x_0}\lt(\beta-\f{\alpha}{u(1+\varphi'(u))}\rt)_+\f{1+\varphi'(u)}{\alpha\varphi(u)}\, du\\
&=&\f\beta\alpha\int_{x_\iy}^{x_0}\lt(\f{1+\varphi'(u)}{\varphi(u)}-\f{\alpha}{\beta}\f1{u\varphi(u)}\rt)_+\, du
=\cJ(\varphi)\eq
\wwtbp
\par
Let us now extend the above transformation to a policy $b\in\cB$ which can take the value 0.
More precisely consider two times $0\leq t_1\leq t_2$ such that $[t_1,t_2]$  or $(t_1,t_2]$ is a connected component of the set $\{t\geq 0\st b(t)=0\}$.
 We recall that this set is assumed to be finite union of such intervals, if it is not empty.\par
Let us first suppose that $t_1\neq t_2$. On $(t_1,t_2]$, \eqref{ODE3} is transformed into
\bq
\lt\{
\begin{array}{rcl}
\dot{x}(t)&=&0\\[2mm]
 \dot{y}(t)&=&-\alpha y(t)
\end{array}
\rt.\eq
namely
\bq
\fo t\in(t_1,t_2],\qquad
x(t)&=&x(t_1)\\
y(t)&=&y(t_1)\exp(-\alpha (t-t_1))\eq
\par
Since $x$ remains constant and $y$ is changing, one cannot represent $y$ as a function $\varphi$ of $x$.
To circumvent this difficulty, we allow $\varphi$ to jump at $x(t_1)$, taking
\bq
\varphi(x(t_1))&\df&y(t_1)\\
\varphi(x(t_1)-)&\df& y(t_2)\ =\ y(t_1)\exp(-\alpha (t_2-t_1))\eq
\par
This is illustrated in the following diagram:
\nopagebreak
 \begin{center}
  \includegraphics[width=9cm]{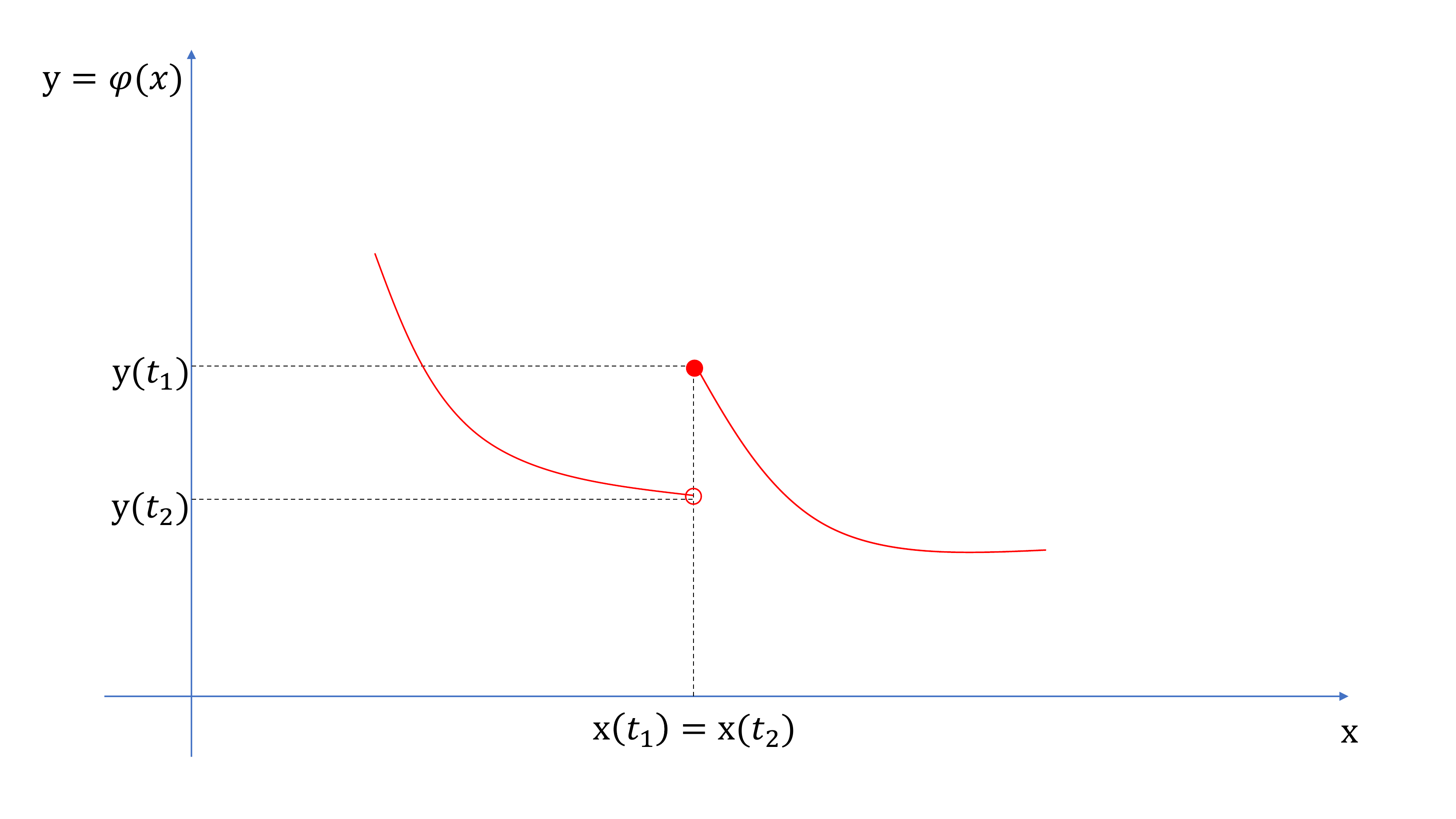}
\end{center}
\par
The contribution of the period $(t_1,t_2]$ to the cost $\cC(b)$ is
\bq
\int_{t_1}^{t_2} (\beta-0)_+\, dt&=&\beta (t_2-t_1)=\f\beta\alpha \ln\lt(\f{y(t_1)}{y(t_2)}\rt)
=\f\beta\alpha \ln\lt(\f{\varphi(x(t_1)}{\varphi(x(t_1)-)}\rt)
\eq
\par
The above observations are also valid, but trivial, when $t_2=t_1$.\par
A priori, our definition of $\cB$ does not exclude the fact that $t_2=+\iy$, namely that $b$ ends up vanishing identically after $t_1$.
By convention in this case $\varphi(x(t_1)-)=0$ and the above formula gives an infinite contribution to the cost, which is
coherent with the fact that
$\int_{t_1}^{+\iy} (\beta-b(t))_+\, dt=\int_{t_1}^{+\iy} \beta\, dt=+\iy$. Since this situation is not interesting for our optimization problem,
we exclude it from our considerations and from now on the connected components of $\{b=0\}$ are assumed to be bounded.
\par
\sm
Repeating the above treatment to all the connected components of $\{b=0\}$ and extending Lemmas~\ref{cB+} and \ref{cB+2}
to the connected components of $\{b>0\}$,
we can associate to any $b\in\cB$ a function $\varphi$ satisfying
\begin{itemize}
\item[(H1):] $\varphi$ is defined on $(x_\iy,x_0]$, takes values in $(0,1)$ and $\varphi(x_0)=y_0$,
\item[(H2):] $\varphi$ has at most a finite number of discontinuity points and is right continuous and admits a positive left limit at them,
\item[(H3):] at any discontinuity point $u$, $ \varphi(u)>\varphi(u-)$,
\item[(H4):] $\varphi$ admits a right derivative $\varphi'$, as well as a left derivative,
outside a finite number of points (which includes the discontinuity points of $\varphi$, but also some points
corresponding to the case $t_2=t_1$ described above, since due to  \eqref{ppp}, at these points $\varphi'$ may diverge to $+\iy$),
\item[(H5):] $\varphi'>  -1$ where it is defined,
\end{itemize}
\par
But the most important feature is that
\bq
\cC(b)&=& \cJ(\varphi)\eq
where the functional is defined by
\bq
\cJ(\varphi)&\df& \int_{(x_\iy, x_0)} L(\xi,\varphi(\xi),\varphi'(\xi))\, d\xi
+\f{\beta}{\alpha}\sum_{u\in (x_\iy, x_0]\st \varphi(u)\neq\varphi(u-)}\!\!\! \ln\lt(\f{\varphi(u)}{\varphi(u-)}\rt)
\eq
($x_\iy$ is excluded from the last sum, since we don't allow $b$ to end up vanishing identically).
\par
There is a point which is not satisfactory in the above construction of $\varphi$: its definition domain $(x_\iy,x_0]$  still depends explicitly on $b$ via $x_\iy$.
It is possible to erase this drawback with the convention that $\varphi$ vanishes on $[0,x_\iy]$
Accordingly, $L$ has to be extended to $[0,x_0]\times [0,1)\times (-1,+\iy)$ via the convention that
$L(\xi, \chi,\chi')=0$ if $\xi\in[0,x_\iy]$ (or equivalently if $\chi=0$).
\par
Remarking  that the condition $y\leq \gamma$ can be translated into $\varphi\leq \gamma$, 
we have embedded the problem of the global minimization of $\cC$ over $\cB_{\gamma}$ into the  problem of the global minimization of $\cJ$ over $\wi\cF_{\gamma}$, the set of
functions $\varphi$ that satisfy the  requirements (H1a), (H2), (H3), (H4) and (H5), where (H1) has been replaced by
\begin{itemize}
\item[(H1a):] $\varphi$ is defined on $[0,x_0]$, takes values in $[0,\gamma]$,
$\varphi(0)=0$, $\varphi(x_0)=y_0$ and if $\varphi(u)>0$ for some $u\in(0, x_0]$, then $\varphi(v)>0$  and $\varphi(v-)>0$ for all $v\in[u,x_0]$.
\end{itemize}
\par
As it was explained in the main text,
the most important part of the above optimization problem
concerns the contribution of the restriction of the function $\varphi$ to the interval $[\alpha/\beta, x_0]$,
since once $x$ has reached $\alpha/\beta$, the \textit{laissez-faire} policy is cost-free and induces $y$ to be non-increasing.
Let us furthermore assume the function $\varphi$ satisfies 
\bqn{abg}
\varphi(\alpha/\beta)&=&\gamma\eqn
\par
This restriction will be justified a posteriori (see Subsection \ref{btt}), but it can also be understood a priori according to the following heuristic.
Let us come back to the temporal description given by $(b,x,y)$.
To get $x_\iy\geq \alpha/\beta$ is extremely costly and requires in fact $\cC(b)=+\iy$
(think for instance to the case where $x_\iy=x_0$ which asks that $b\equiv 0$), so for the optimization problem at hand, we can dismiss this
possibility and assume $x_\iy<\alpha/\beta$.
Then consider $\tau_+$ the first time $x$ reaches $\alpha/\beta$, we have $y(\tau_+)\in (0,\gamma]$.
As already observed, taking $b=\beta$ after time $\tau_+$ leads to a  non-increasing evolution of $y$  on $[\tau_+,+\iy)$.
In particular, $y(t)$ stays below $\gamma$ for all times $t\in [\tau_+,+\iy)$, while this part of the trajectory does not participate positively 
to the cost.
Due to our underlying assumption, the \textit{laissez-faire} policy $b\equiv\beta$
leads $y$ to hit $\gamma$ strictly before $x$ reaches $\alpha/\beta$.
Then if $y(\tau_+)<\gamma$, it means that for some times $t\in[0,\tau_+)$, we have $b(t)<\beta$ and $y(t)<\gamma$.
Increasing a little $b$ at such times, will have the effect of increasing $y(\tau_+)$ while keeping $y$ below $\gamma$.
Furthermore this operation will decrease a little the cost.
As a consequence, in order to find a minimizing policy $b$ for $\cC$, we can assume that
$y(\tau_+)=\gamma$. This amounts to assuming \eqref{abg}.
\par\sm
These considerations and Assumption \eqref{abg} lead us to modify the functional set $\wi\cF_{\gamma}$ into $\cF_{\gamma}$,
 replacing its first requirement (H1a) by
\begin{itemize}
\item[(H1b):] $\varphi$ is defined on $[\alpha/\beta, x_0]$, takes values in $(0,\gamma]$, the left limits of $\varphi$ are positive,
$\varphi(\alpha/\beta)=\gamma$ and $\varphi(x_0)=y_0$
\end{itemize}
\par
Of course the cost functional on such functions $\varphi$ is given by
\bqn{cJ2}
\nonumber\cJ(\varphi)&\df& \int_{(\alpha/\beta, x_0)} L(\xi,\varphi(\xi),\varphi'(\xi))\, d\xi
\\&&+\f{\beta}{\alpha}\sum_{u\in (\alpha/\beta, x_0]\st \varphi(u)\neq\varphi(u-)} \ln\lt(\f{\varphi(u)}{\varphi(u-)}\rt)
\eqn
\par
Our ultimate goal is to prove that $\cJ$ admits a unique minimizer $\varphi^*$ over $\cF_{\gamma}$ and that this
minimizer is obtained from $b^*$ by the operation described above. Up to the justification of the reduction of (H1a) to (H1b), given at the end of Subsection \ref{btt},
Theorem \ref{Thm: Main} will then be proven.

\subsection{Extension of $\cJ$ to measures}\label{eoJtm}

To prove the existence of the minimizer $\varphi^*$ of $\cJ$ on $\cF_{\gamma}$, we begin by generalizing this optimization problem by replacing $\cF_{\gamma}$ by a set of measures.\par\sm
To any given $\varphi\in\cF_{\gamma}$, we associate three measures $\mu$, $\psi$ and $\nu$ on 
$I\df[\alpha/\beta, x_0]$
via
\bq
\mu(dx)&\df& \f{\varphi'(x)}{\varphi(x)}\,dx+\sum_{u\in (\alpha/\beta, x_0]\st \varphi(u)\neq\varphi(u-)} \ln\lt(\f{\varphi(u)}{\varphi(u-)}\rt)\delta_u(dx)\\
\psi(dx)&\df& \f1{\varphi(x)}\,dx\\
\nu(dx)&\df& \f1{\varphi(x)}\lt(1-\f{\alpha}{\beta x}\rt)\,dx\eq
(where $\delta_{u}$ stands for the Dirac mass at $u$).
\par
Note that $\psi$ and $\nu$ are non-negative measures, but $\mu$ is a signed measure.
Denote $F_{\mu}$ the repartition function associated to $\mu$ via
\bq
\fo x\in I,\qquad F_{\mu}(x)&\df &\mu([\alpha/\beta, x])\eq
\par
Recall that $\varphi$, as well as its left limits, are positive on $I$, it follows that $\varphi$ is bounded below by a positive constant on $I$.
This observation enables us to compute $F_{\mu}$:
\bqn{Fmu}
\fo x\in I,\qquad F_{\mu}(x)&=&\ln(\varphi(x))-\ln(\varphi(\alpha/\beta))=
\ln(\varphi(x)/\gamma)\eqn
\par
We deduce that
\bq\fo x\in I,\qquad\varphi(x)&=&\gamma \exp(F_{\mu}(x))\eq
and as a consequence
\bq
\psi(dx)&\df& \f{\exp(-F_{\mu}(x))}{\gamma}dx\\
\nu(dx)&\df& \lt(1-\f{\alpha}{\beta x}\rt)\f{\exp(-F_{\mu}(x))}{\gamma}dx\eq\par
On these expressions, it appears that $\psi$ and $\nu$ only depend on $\mu$ (in addition to the constants $\alpha,\beta,\gamma$), so they will be denoted $\psi_{\mu}$
$\nu_{\mu}$ from now on.
\par
From \eqref{Fmu} we get that the total weight $\mu(I)$ of $\mu$ is given by $F_{\mu}x_0=\ln(y_0/\gamma)$.
Let us estimate the total variation $\lVe \mu\rVtv$ of $\mu$:
\begin{lem}\label{tv}
We have
\bq
\lVe \mu\rVtv&\leq &\f2{\epsilon}(x_0-\alpha/\beta)+\ln(y_0/\gamma)\eq
where 
\bq
\epsilon&\df& \inf\{\varphi(x)\st x\in I\}\ =\ \min\{\varphi(x)\wedge \varphi(x-)\st x\in I\}\eq
\end{lem}
\prooff
Recall that any signed measure $m$ on $I$ can be decomposed into $m_+-m_-$, where $m_-$ and $m_+$ are two non-negative measures mutually singular.
The total variation is given by $\lVe m\rVtv= m_-(I)+m_+(I)$.
\par
Coming back to $\mu$, we have
\bq
\mu_-(dx)&=&\f{\lve \varphi'(x)\rve}{\varphi(x)}\un_{\{\varphi'(x)>0\}}\, dx\eq
so that
\bq
\mu_-(I)&\leq & \int_I \f1{\varphi(x)}\, dx
\leq  \f1\epsilon \int_I 1\, dx
=\f{x_0-\alpha/\beta}{\epsilon}
\eq
\par It follows that
\bq
\lVe \mu\rVtv&=&\mu_-(I)+\mu_+(I)
=2\mu_-(I)+\mu(I)
\leq 
\f2{\epsilon}(x_0-\alpha/\beta)+\ln(y_0/\gamma)\eq
\wwtbp
\par
The quantity $\epsilon>0$ associated to $\varphi$ in the previous lemma can be estimated in terms of $\cJ(\varphi)$:
\begin{lem}\label{tv2}
We have
\bq
\epsilon&\geq & y_0\exp(-\beta \cJ(\varphi)/\alpha)\eq
\end{lem}
Fix an arbitrary point $x\in I$.
We have
\bq
\cJ(\varphi)&\geq & \f\beta\alpha\int_{I} L(u,\varphi(u),\varphi'(u))\, du
\geq  \f\beta\alpha\int_x^{x_0} L(u,\varphi(u),\varphi'(u))\, du\\
 &=&\f\beta\alpha\int_x^{x_0}\lt(\f{\varphi'(u)}{\varphi(u)}+\f{\alpha}{\beta\varphi(u)}\lt(1-\f{\alpha}{\beta u}\rt)\rt)_+\, du\\
&\geq & \f\beta\alpha\int_x^{x_0}\f{\varphi'(u)}{\varphi(u)}+\f{\alpha}{\beta\varphi(u)}\lt(1-\f{\alpha}{\beta u}\rt)\, du\\
&\geq & \f\beta\alpha\int_x^{x_0}\f{\varphi'(u)}{\varphi(u)}\, du=\f\beta\alpha\ln\lt(\f{\varphi(x_0)}{\varphi(x)}\rt)
\eq
where in the last-but-one inequality, we took into account that $1-\f{\alpha}{\beta u}\geq 0$ for $u\geq x\geq \alpha/\beta$.
The above bound can be written
\bq
\varphi(x)&\geq &\varphi(x_0)\exp(-\alpha \cJ(\varphi)/\beta)
=y_0\exp(-\alpha \cJ(\varphi)/\beta)\eq
\par
By taking the infimum over all $x\in I$, we get the desired result.
\wwtbp
\par
Consider an element $\varphi_0$ of $\cF_{\gamma}$ such that $\cJ(\varphi_0)<+\iy$, for instance the function constructed from $b^*$
as in the previous subsection and denote
\bqn{M}
M&\df& \f{2\exp(\alpha (1+\cJ(\varphi_0))/\beta)}{y_0}(x_0-\alpha/\beta)+\ln(y_0/\gamma)\eqn
\par
The global minimization of $\cJ$ on $\cF_{\gamma}$ is equivalent of the global minimization
of $\cJ$ on $\{\varphi\in\cF_{\gamma}\st \cJ(\varphi)\leq \cJ(\varphi_0)\}$.
So according to Lemma~\ref{tv}, we can restrict our attention to measure $\mu$
satisfying
\begin{itemize}
\item[(C1):] $\lVe\mu\rVtv\leq M$.
\end{itemize}
Furthermore note that the belonging of $\varphi$ to $\cF_{\gamma}$ implies three conditions on $\mu$:
\begin{itemize}
\item[(C2):] $\mu(I)=\ln(y_0/\gamma)$,
\item[(C3):] $F_{\mu}\leq 0$,
\item[(C4):] $\mu+\psi_{\mu}\geq 0$.
\end{itemize}
\par
Denote by $\cM_{\gamma}$ the set of signed measures $\mu$ on $I$ which satisfy the conditions (C1), (C2), (C3) and (C4).
\par
Up to now, we did not use $\nu_{\mu}$, its interest comes from the fact that 
for $\varphi\in \cF_{\gamma}$, the cost functional writes
\bq
\cJ(\varphi)&=&\f{\alpha}{\beta}(\mu+\nu_{\mu})_+(I)\eq
\par
This observation leads us to define for any $\mu\in \cM_{\gamma}$,
\bq
\cK(\mu)&=&\f{\alpha}{\beta}(\mu+\nu_{\mu})_+(I)\eq
since the global minimization of $\cJ$ on $\{\varphi\in\cF_{\gamma}\st \cJ(\varphi)\leq \cJ(\varphi_0)\}$ can be embedded
in the global minimization of $\cK$ on $\cM_{\gamma}$.

\subsection{Existence of a global minimizer of $\cK$ on $\cM_{\gamma}$}

The successive extensions of our initial minimization problem worked out in the two previous subsections
will be justified here by showing the existence of a global minimizer, via abstract topological arguments. It won't be clear that such a minimizer from $\cM_{\gamma}$
corresponds to an element of $\cB_{\gamma}$ via the transformation of Subsection \ref{txps}: this will be investigated in the next subsections.
\par\sm
Let us endow the set of (signed) measures on $I$ with the weak topology, i.e.\ a sequence $(\mu_n)_{n\in\NN}$ of such measures converges toward
a measure $\mu$ if and only if for any continuous function $g\st I\ri \RR$, we have
\bq
\lim_{n\ri \iy}\mu_n[g]&=&\mu[g]\eq
\par
The existence of a global minimizer of $\cK$ on $\cM_{\gamma}$ is a consequence of a version of Weierstrass' maximum theorem through the two following results.
\begin{pro}\label{pro1}
The set $\cM_{\gamma}$ is compact.
\end{pro}
\par
\begin{pro}\label{pro2}
The mapping $\cK\st \cM_{\gamma}\ri\RR$ is lower semi-continuous.
\end{pro}
\par
\proofff{Proof of Proposition \ref{pro1}}
Consider the ball (with respect to the strong topology) $B(M)$ consisting of the signed measures
whose total variation is smaller or equal to $M$.
It is well-known that $B(M)$ is weakly compact.
So it sufficient to show that
the sets
\bq
S_1&\df&\{\mu\in B(M)\st \mu(I)=\ln(y_0/\gamma)\}\\
S_2&\df&\{\mu\in B(M)\st F_{\mu}\leq 0\}\\
S_3&\df&\{\mu\in B(M)\st \mu+\psi_{\mu}\geq0\}\eq
are closed.\par
$\bullet$ Concerning $S_1$, this is obvious, since $\mu(I)=\mu[\un_I]$, where $\un_I$ is the continuous function on $I$
always taking the value 1.\par
$\bullet$ For $S_2$, consider a sequence $(\mu_n)_{n\in\NN}$ of measures from $S_2$ converging toward a signed measure $\mu$.
We have to check that $\mu\in S_2$.
Consider $\cA$ the set of atoms of $\mu$, $\cA$ is at most denumerable and for $x\in I\setminus \cA$, we have
\bqn{dense}
\lim_{n\ri\iy} F_{\mu_n}(x)&=&F_{\mu}\eqn
\par
We deduce
\bq
\fo x\in I\setminus \cA,\qquad F_{\mu}(x)&\leq &0\eq
It remains to use that $F_{\mu}$ is right-continuous and that $I\setminus \cA$ is dense in $I$ to extend 
the validity of the above inequality to the whole set $I$.\par
$\bullet$ For $S_3$, consider a sequence $(\mu_n)_{n\in\NN}$ of measures from $S_3$ converging toward a signed measure $\mu$.
We have to check that $\mu+\psi_{\mu}\geq 0$.
Consider $g$ a continuous function on $I$. We begin by showing that
\bqn{convpsi}
\lim_{n\ri \iy}\psi_{\mu_n}[g]&=&\psi_{\mu}[g]\eqn
Indeed, for any $n\in\NN$, we have
\bq
\psi_{\mu_n}[g]&=&\f1\gamma\int_I g(x) \exp(-F_{\mu_n}(x))\, dx\eq\par
We have seen in \eqref{dense} that $F_{\mu_n}$ is almost everywhere converging to $F_{\mu}$.
Furthermore, we have
for all $n\in\NN$, 
\bq
\fo x\in I,\qquad \vert F_{\mu_n}(x)\vert &\leq & \lVe \mu_n\rVtv
\leq  M\eq\par
It follows that dominated convergence can be invoked to conclude to \eqref{convpsi}.
\par
We deduce that
\bq
\lim_{n\ri\iy}(\mu_n+\psi_{\mu_n})[g]&=&(\mu+\psi_{\mu})[g]\eq
\par
Now assume furthermore that $g\geq 0$. The above convergence implies that
\bq
(\mu+\psi_{\mu})[g]&\geq &0\eq
Since this is true for all non-negative continuous function $g$, we get that $\mu+\psi_{\mu}\geq 0$, as desired.
\wwtbp
\par
\proofff{Proof of Proposition \ref{pro2}}
To get that the mapping $\cK\st \cM_{\gamma}\ri\RR$ is lower semi-continuous, it is sufficient to write it as
the supremum of (weakly) continuous functions on $\cM_{\gamma}$.
By definition, we have
\bq
\cK(\mu)&=&\f{\alpha}{\beta}(\mu+\nu_{\mu})_+(I)
=\sup_{g\in\cC(I)\st 0\leq g\leq 1}(\mu+\nu_{\mu})[g]
\eq
where $\cC(I)$ is the set of continuous functions on $I$.
\par
So it remains to check that for any fixed $g\in\cC(I)$ with $0\leq g\leq 1$,
the mapping
\bq
\cM_{\gamma}\ni \mu&\mapsto& (\mu+\nu_{\mu})[g]\eq
is continuous. 
The proof of this continuity is similar to the closure of $S_3$ in the proof of Proposition~\ref{pro1}:
both the mappings $\cM_{\gamma}\ni \mu\mapsto \psi_{\mu}$ 
and $\cM_{\gamma}\ni \mu\mapsto \nu_{\mu}$ are continuous (taking values in the set of non-negative measures on $I$ endowed with the weak topology).
\wwtbp

\subsection{Reduction to absolutely measures}\label{rtam}

Let $\mu^*$ be a minimizer of $\cK$ on $\cM_{\gamma}$. We will show here that $\mu^*$ is absolutely continuous with respect to $\lambda$, the Lebesgue measure on $I$.
\par\sm
Recall that any signed measure $\mu$ on $I$ can be uniquely decomposed into a sum $\mu_{\mathrm{a}}+\mu_s+\mu_{\mathrm{c}}$, where $\mu_{\mathrm{a}}$ is atomic, $\mu_{\mathrm{c}}$ is diffuse and singular with respect to $\lambda$
and $\mu_{\mathrm{c}}$ is absolutely continuous with respect to $\lambda$. Due to Radon-Nikodym theorem,  $\mu_{\mathrm{c}}$ admits a (signed) density $f\st I\ri\RR$ with respect to $\lambda$,
this relation will be denoted $\mu_{\mathrm{c}}=f\cdot \lambda$.
\par
Using this decomposition, the cost functional $\cK$ can be written under the form
\bq
\cK(\mu)&=&\f{\beta}{\alpha}\lt(\mu_{\mathrm{a}}(I)+\mu_s(I)+\int_I (f+\nu_{\mu})_+\, d\lambda\rt)\eq
where we identified $\nu_{\mu}$ with its density with respect to $\lambda$, namely
\bq
\fo x\in I,\qquad \nu_{\mu}(x)&=&\f1{\gamma}\lt(1-\f{\alpha}{\beta x}\rt)\exp(-F_{\mu}(x))\eq
\par
Our goal here is to show that $\mu^*_{\mathrm{a}}=0=\muss$, by perturbative arguments. Both proofs will follow the same pattern, but the deduction of $\mu_{\mathrm{a}}=0$
is simpler, so for pedagogical reasons we will insist on this one.
\par
\proofff{Proof of $\mu^*_{\mathrm{a}}=0$}
The argument is by contradiction. Assume that $\mu^*_{\mathrm{a}}\neq 0$, then there exist $x_1\in I$ and $\epsilon>0$ such that
$\mu^*_{\mathrm{a}}\geq \epsilon\delta_{x_1}$.
Since 
$F_{\mus}(x_1)-F_{\mus}(x_1-)\geq \epsilon$ and that $F_{\mus}(x_1)\leq 0$,
we have $F_{\mus}(x_1-)\leq -\epsilon$ and we can find $x_2\in(\alpha/\beta,y_1)$ such that
\bq
\fo x\in[x_2,x_1),\qquad F_{\mus}(x)&\leq&-\f{\epsilon}2\eq
\par
It leads us to consider for $r> 0$, the perturbation
$\musr$ defined by
\bq
\musr&\df& \mu+r\f{\un_{[x_2,x_1)}}{x_1-x_2}\cdot \lambda-r\delta_{x_1}\eq
(where $\un_{[x_2,x_1)}$ is the indicator function of $[x_2,x_1)$ and the middle term
is an absolutely continuous measure).
\par
Let us check that $\mu_r\in \cM_{\gamma}$ for $r> 0$ small enough, namely that (C1), (C2), (C3) and (C4) are satisfied by $\musr$.
\par\sm
$\bullet$ $\lVe \musr\rVtv\leq M$:
\par
Since $\mus$ is also a minimizer of $\cK$ on the space of signed measures on $I$,
we have $\cK(\mus)\leq \cL(\varphi_0)$, where $\varphi_0$ was defined above \eqref{M}.
It follows from Lemmas \ref{tv} and \ref{tv2}, taking into account the definition of $M$ in \eqref{M},
that $\lVe \mus\rVtv< M$.
Triangular inequality with respect to the total variation norm implies
\bq
\lVe \musr\rVtv&\leq & \lVe \mus\rVtv+\lVe r\f{\un_{[x_2,x_1)}}{x_1-x_2}\cdot \lambda-r\delta_{x_1}\rVtv
= \lVe \mus\rVtv+ 2r\eq
insuring that for $r\leq  (M-\lVe \mus\rVtv)/2$, we have $\lVe \mus\rVtv\leq M$.\par\sm
$\bullet$ $\musr(I)=\ln(y_0/\gamma)$:
\par
The total weight of $\musr$ is always the same as that of $\mus$, since
\bq
\lt(\f{\un_{[x_2,x_1)}}{x_1-x_2}\cdot \lambda-\delta_{x_1}\rt)(I)&=&1-1\ =\ 0\eq
\par\sm
$\bullet$ $F_{\musr}\leq 0$:\par
Note that outside $(x_2,x_1)$,  $F_{\musr}$ coincides with $F_{\mus}$, 
so we just need to check that $ F_{\musr}(x)\leq 0$ for all $x\in(x_1,x_2)$. Indeed, we have
\bq
\fo x\in(x_1,x_2),\qquad F_{\musr}(x)&=&
F_{\mus}(x)+r\f{x-x_2}{x_1-x_2}
\leq  -\f{\epsilon}{2}+r
\leq  0\eq
as soon as $r\leq \epsilon/2$.\par
\sm
$\bullet$ ${\musr}+\psi_{\musr}\geq 0$:\par
Note again that
outside $(x_2,x_1]$, we have ${\musr}+\psi_{\musr}={\mus}+\psi_{\mus}$, 
so we just need to check that the measure  ${\musr}+\psi_{\musr}$ is non-negative on $(x_2,x_1]$. 
The diffusive singular part of ${\musr}+\psi_{\musr}$ is the same as that of ${\mus}+\psi_{\mus}$
and the atomic ones only differ at $x_1$. Note that $\musr(\{x_1\})= \mu_s(\{x_1\})-r\geq \epsilon-r$
and this is non-negative, as soon as $r\leq \epsilon$.
Concerning the absolutely continuous part,
denote $f^*$ (respectively $f^*_r$) the density of $\mus$ (resp.\ $\musr$) with respect to $\lambda$.
We have a.e.\ for $x\in (x_2,x_1)$,
\bqn{fsi}
f^*_r(x)&=&f^*(x)+\f{r}{x_1-x_2}\eqn\par
Identify $\psi_{\musr}$ with its density, so that we can write (a.e.)
\bq
\fo x\in (x_2,x_1), \qquad \psi_{\musr}(x)&=&\f{\exp(-F_{\musr}(x))}{\gamma}\eq
\par
Note that 
\bq
\fo x\in (x_2,x_1), \qquad \lve F_{\musr}(x)-F_{\mus}(x)\rve&=&\lve \int _{x_2}^x\f{r}{x_1-x_2}\,d\lambda\rve
=\f{r(x-x_2)}{x_1-x_2}\ \leq \ r
\eq
Let
 \bq C&\df& \f{\exp(1+\max_{x\in [x_2,x_1]}\lve F_{\mu}(x)\rve)}{\gamma}\eq
\par
For any $r\in[0,1]$, we have
\bq
\fo x\in (x_2,x_1), \qquad
\lve 
\psi_{\musr}(x)-\psi_{\mus}(x)\rve&\leq & Cr\eq
\par
Comparing with \eqref{fsi}, we get that if $x_2$ has been chosen so that $x_1-x_2\leq C^{-1}$,
then a.e.\ for $x\in (x_2,x_1)$,
\bq
f^*_r(x)+\psi_{\musr}(x)\geq f^*(x)+\psi_{\mus}(x)\ \geq \ 0\eq
\par\me
This ends the proof that $\mu_r\in\cM_{\gamma}$, as soon as $r>0$ is small enough and 
 $x_2$ is sufficiently close to $x_1$.\par
Let us evaluate $\cK(\musr)$.
We have
\bqn{cKcK}
\hskip-5mm\cK(\musr)&=&\cK(\mus)-r+\int_{x_2}^{x_1} \lt(f+\f{r}{x_1-x_2}+\nu_{\musr}\rt)_+-(f+\nu_{\mu})_+\, d\lambda\eqn
and for almost all $x$ belonging to $(x_2,x_1)$,  
\bq
\lt(f(x)+\f{r}{x_1-x_2}+\nu_{\musr}(x)\rt)_{\!+}\!\!\!\!-(f(x)+\nu_{\mu}(x))_{+}
\!\leq \! \f{r}{x_1-x_2}+\nu_{\musr}(x)-\nu_{\mu}(x)\eq
Note that for $x\in(x_2,x_1)$, we have $F_{\musr}(x)>F_{\mus}(x)$, so that
$\nu_{\musr}(x)<\nu_{\mus}(x)$. Thus we get
\bq
 \lt(f(x)+\f{r}{x_1-x_2}+\nu_{\musr}(x)\rt)_+-(f(x)+\nu_{\mu}(x))_+&< &\f{r}{x_1-x_2}
\eq
\par
It follows that that for $r>0$:
\bq
\int_{x_2}^{x_1} \lt(f+\f{r}{x_1-x_2}+\nu_{\musr}\rt)_+-(f+\nu_{\mu})_+\, d\lambda&< &r\eq
and \eqref{cKcK} implies the contradiction
$\cK(\musr)<\cK(\mus)$.\wwtbp
\par
\begin{rem} 
In Subsection \ref{eoJtm}, we have seen that the atomic part of a measure corresponds to imposing the drastic policy $b=0$ for some time,
as a partial attempt toward eradication of the disease (according to Remark \ref{eradication} this goal cannot be fully attained).
The significance of $\musa=0$ is that such attempts are sub-optimal. From the above proof we see that it is better to replace such attempts by future softer policies,
replacing a (partial) Dirac mass at $x_1$ by a density before $x_1$ (recall that $x$ and time go in reverse directions).
\end{rem}
\par
\proofff{Proof of $\muss=0$}
The pattern of the proof is the same as that for $\musa=0$, except that we have to ``thicken a little'' $x_1$.
Indeed, if $\muss\neq 0$, we can find $x_1\in(\alpha/\beta, x_0]$ and $\epsilon'>0$ such that
\bq
\f{\alpha}{\beta}&<& x_1-\epsilon'\\
\muss((x_1-\epsilon',x_1+\epsilon'))&>&0\\
\fo \epsilon''\in(0,\epsilon'],\qquad \muss((x_1-\epsilon'',x_1+\epsilon''))&\geq & 2\musc((x_1-\epsilon'',x_1+\epsilon''))\eq
\par
In comparison with the previous proof, the restriction of $\muss$ to $(x_1-\epsilon',x_1+\epsilon')$ plays the role of
$\epsilon \delta_{x_1}$ with $\muss((x_1-\epsilon',x_1+\epsilon'))$ playing the role of $\epsilon$.\par
It is now possible to find $x_2\in(\alpha/\beta,x_1-\epsilon')$ such that for $r>0$ small enough,
the measure 
\bq\musr&\df& \mu+r\f{\un_{[x_2,x_1-\epsilon')}}{x_1-\epsilon'-x_2}\cdot \lambda-\f{r}{\muss([x_1-\epsilon',x_1+\epsilon'])}\un_{(x_1-\epsilon',x_1+\epsilon')}\cdot\muss\eq
belongs to $\cM_{\gamma}$ and
\bq\cK(\musr)&<&\cK(\mus)\eq
\par
This is the desired contradiction. The details of the adaptation of the arguments of the proof of $\musa=0$ are left to the reader.\wwtbp
\par
We have shown that any minimizer $\mu^*$ of $\cK$ can be written in the form $f^*\cdot\lambda$.
As a consequence, such a density $f^*$ is a minimizer of a suitably formulated optimization problem, to which we now turn..
\par
For any integrable function $f$ on $I$, let associate to $f$ the notions that were previously associated to the measure $f\cdot\lambda$: for any $x\in I$,
\bq
F(x)&\df & F_{f\cdot \lambda}(x)\ =\ \int_{\alpha/\beta}^xf(u)\, du\\
\psi_F(x)&\df& \f1\gamma\exp( -F(x))\\
\nu_F(x)&\df&\lt(1-\f\alpha{\beta x}\rt)\exp(-F(x))
\eq
\par
We introduce
$\cD_{\gamma}$ the set of integrable functions $f$ such that
\bq
F(x_0)=\ln(y_0/\gamma),\quad
F\leq  0,\quad
f+\psi_F\geq 0\eq
and we consider on $\cD_{\gamma}$ the functional
\bq
\cG(f)\df\cK(f\cdot\lambda)
=\int_I(f+\nu_F)_+\, d\lambda\eq
\par
We have shown the optimization problem of $\cG$ on $\cD_{\gamma}$ (which is an extension of the optimization
problem of $\cJ$ on $\cF_{\gamma}$) admits global minimizers: they are exactly the functions $f^*$ such that
$f^*\cdot\lambda\in\cM_{\gamma}$ is a global minimizer of $\cK$.

\subsection{Characterization of the minimizer of $\cG$ on $\cD_{\gamma}$}\label{cotmoGoD}

In this subsection, a minimizer of $\cG$ on $\cD_{\gamma}$ will be denoted $f$ and our goal is to show
that it is a.e.\ equal to the function $f^*$ described below.
\par\sm
Define $x^*$ as the unique solution belonging to $[\alpha/\beta, x_0]$ of the equation
\bqn{xs}
x^*-\f{\alpha}{\beta}\ln(x^*)&=&1-\gamma-\f{\alpha}{\beta}\ln(x_0)\eqn
(the existence of this solution is due to our underlying assumption, note that $x^*$ coincides with $x(\tau_1)$ defined above \eqref{x}),
and take
\bqn{fs2}
\nonumber\lefteqn{\hskip-10mm\fo x\in[\alpha/\beta, x_0],}\\
\hskip-5mmf^*(x)&\df& \lt\{ \begin{array}{ll}
0&\hbox{, if $x\leq x^*$}\\
\!-\lt(\gamma -\lt(x-x^*-\f\alpha{\beta }\ln(x/x^*)\rt)\rt)^{\!-1}\!\!\lt(1-\f\alpha{\beta x}\rt)&\hbox{, if $x>x^*$}
\end{array}\rt.
\eqn
\par
\begin{theorem}\label{theofs}
The function $f^*$ is the unique minimizer of $\cG$ on $\cD_{\gamma}$, up to modifications on  subsets of $I$ with Lebesque measure zero.
\end{theorem}
\par
The proof of this result  proceeds by way of a few lemmas.
\par
We associate to $f$, a minimizer of $\cG$ on $\cD_{\gamma}$, the notions defined at the end of the previous subsection.
In addition, we
define
\bq
\z&\df& \max\{x\in I\st F(x)=0\}\\
\cG_1(f)&\df&\int_{\alpha/\beta}^{\z} (f+\nu_F)_+\, d\lambda\\
\cG_2(f)&\df&\int_{\z}^{x_0} (f+\nu_F)_+\, d\lambda\eq
\par
Note that $\cG(f)=\cG_1(f)+\cG_2(f)$. We will investigate separately $\cG_1(f)$ and $\cG_2(f)$.
The computation of the former will be a consequence of:
\begin{lem}\label{l1}
The function $F$ vanishes everywhere on $[\alpha/\beta, \z]$.
\end{lem}
\prooff
We have
\bq
\cG_1(f)&\geq & \int_{\alpha/\beta}^{\z} f+\nu_F\, d\lambda
=F(\z)+ \int_{\alpha/\beta}^{\z}\nu_F\, d\lambda
= \int_{\alpha/\beta}^{\z}\nu_F\, d\lambda\eq
since by continuity of $F$, $F(\z)=0$.
\par
The bound $F\leq 0$ implies that
\bqn{nuF}
\fo x\in I,\qquad \nu_F(x)
&=&\f1\gamma\lt(1-\f{\alpha}{\beta x}\rt)\exp(-F(x))
\geq \f1\gamma\lt(1-\f{\alpha}{\beta x}\rt)\eqn
and we deduce
\bq
\cG_1(f)&\geq &  \f1\gamma\int_{\alpha/\beta}^{\z}1-\f{\alpha}{\beta x}\, dx
=\cG_1(\wi f)
\eq
where the function $\wi f$ is given by
\bq
\fo x\in I,\qquad 
\wi f(x)&\df& \lt\{\begin{array}{ll}
0&\hbox{, if $x\leq \z$}\\
f(x)&\hbox{, if $x>\z$}\end{array}\rt.\eq
\par
Since $F(\z)=0$, we get that for any $x\in[\z,x_0]$,
that
\bq
 \wi F(x)&\df& \int_{\alpha/\beta}^x \wi f(u)\,du
=\int_{\z}^x \wi f(u)\,du
=\int_{\z}^x  f(u)\,du
=F(x)\eq
\par
We deduce that $\wi f+\nu_{\wi F}$ coincides  with 
$f+\nu_F$ on $[\z,x_0]$  and thus $\cG_2(\wi f)=\cG_2(f)$, which implies 
\bq
\cG(\wi f)&\leq & \cG(f)\eq
\par
Taking into account that $\wi f\in\cD_{\gamma}$ and that $f$ is a minimizer of $\cG$ on $\cD_{\gamma}$, we must have $
\cG(\wi f)= \cG(f)$ and $\cG_1(\wi f)=\cG_1(f)$.
In particular, \eqref{nuF} must be an equality a.e.\ on $[\alpha/\beta,\z]$. 
This means that $F$ vanishes a.e.\ on $[\alpha/\beta,\z]$. The continuity of $F$ then implies that $F$ vanishes identically on  $[\alpha/\beta,\z]$.
\wwtbp
\par
As announced, we deduce the value of $\cG_1(f)$:
\begin{lem}\label{cG1}
\bq
\cG_1(f)&=&\f1\gamma\lt(\z-\f\alpha\beta\ln(\z)\rt)+\f\alpha{\beta\gamma}\lt(\ln\lt(\f\alpha\beta\rt)-1\rt)\eq
and the r.h.s.\ is increasing with respect to $\z$.
\end{lem}
\prooff
From Lemma \ref{l1}, we deduce that $f=0$ on $[\alpha/\beta,\z]$ (a.e., as all statements about densities, in the sequel we will no longer mention it).
Recalling that $\nu_F\geq 0$, we get
\bq
\cG_1(f)&=&\int_{\alpha/\beta}^{\z} \nu_F\, d\lambda
=\int_{\alpha/\beta}^{\z} \nu_0\, d\lambda
= \f1\gamma\int_{\alpha/\beta}^{\z}1-\f{\alpha}{\beta x}\, dx\\
&=& \f1\gamma\lt[x-\f{\alpha}{\beta}\ln(x)\rt]_{\alpha/\beta}^{\z}
=
\f1\gamma\lt(\z-\f\alpha\beta\ln(\z)\rt)+\f\alpha{\beta\gamma}\lt(\ln\lt(\f\alpha\beta\rt)-1\rt)\eq
\wwtbp
\par
We now come to the study of $\cG_2$. Our main step in this direction is:
\begin{pro}\label{pro0}
We have $f+\nu_F\leq 0$ on $[\z, x_0]$.
\end{pro}
\prooff
Consider $x_1\in (\z, x_0]$. Since $x_1$ can be arbitrary close to $\z$, it is sufficient to show that $f+\nu_F\leq 0$ on $[x_1,x_0]$.
The advantage of considering such a $x_1$ is that we can find $\eta>0$ such that
\bqn{eta}
\fo x\in[x_1, x_0],\qquad F(x)&\leq &-\eta\eqn
\par
This property will be important for the perturbations of $f$ we are to consider. More precisely they
will be of the form
$f_r\df f+r g$, with $r >0$ sufficiently small and where $g$ is an appropriate bounded function on $[\alpha/\beta, x_0]$
and satisfying $g=0$ on $[\alpha/\beta, x_1]$.
\par
Defining for $x\in[\alpha/\beta, x_0]$
\bq
F_r(x)&\df&
\int_{\alpha/\beta}^x f_r\, d\lambda\\
G(x)&\df& \int_{\alpha/\beta}^x g\, d\lambda\ =\ \int_{\z}^{\z\vee x} g\, d\lambda\eq
we have $F_r=F+r G$ and it is clear from \eqref{eta} that the inequality $F_r \leq 0$ will be satisfied on $[\z, x_0]$
for $r >0$ sufficiently small. Since $F_r$ and $F$ coincide on $[\alpha/\beta, \z]$, we will get $F_r\leq 0$ on $I$.
\par
It will be easy to impose that $G(x_0)=0$, to get $F_r(x_0)=F(x_0)=\ln(y_0/\gamma)$.\par
It will be more tricky to insure that $f_r+\psi_{F_r}\geq 0$ (on $[x_1,x_0]$, since it is trivial on $[\alpha/\beta,x_1]$ where $f_r+\psi_{F_r}$ coincides with
$f+\psi_{F}$)
and we have to be very careful about the choice of $g$. 
The construction of such appropriate $g$ is given as follows.
\par
First note that if $x\in[x_1,x_0]$ is such that $f(x)+\psi_F(x)=0$,
then
\bq
f(x)+\nu_F(x)&=&f(x)+\psi_F(x)-\f{\alpha}{\beta\gamma x}\exp(-F(x))
\leq  -\epsilon\eq
with
\bq
\epsilon&\df& \min\lt\{\f{\alpha}{\beta\gamma x}\exp(-F(x))\st x\in[x_1, x_0]\rt\}\eq
\par
Denote
\bq
A&\df& \{x\in[x_1,x_0]\st f(x)+\nu_F(x)\leq -\epsilon/2\}\\
B&\df& \{x\in[x_1,x_0]\st f(x)+\nu_F(x)>0\}\eq
\par
Let $h$ be a bounded and 
measurable
function defined on $B$.
Consider the two functions $\xi$ and $\chi$ given on $[x_1, x_0]$ by
\bq
\xi (x)&\df&\lt\{\begin{array}{ll}
1&\hbox{, if $x\in A$}\\
h(x)&\hbox{, if $x\in B$}\\
0&\hbox{, otherwise}\end{array}\rt.\\
\chi (x)&\df&\lt\{\begin{array}{ll}
\psi_F(x)&\hbox{, if $x\in A$}\\
\nu_F(x)
&\hbox{, otherwise}\end{array}\rt.
\eq
\par
Solve on $[x_1, x_0]$ the weak ODE in $G$:
\bqn{G}
\lt\{\begin{array}{rcl}
G(x_1)&=&0\\[2mm]
\dot{G}&=&\chi G+\xi
\end{array}\rt.
\eqn
(this is always possible, even with irregular $\xi$ and $\chi$, see \eqref{jG} below).
\par
Next extend $G$ to $[\alpha/\beta, x_1]$ by imposing that $G$ vanishes there (equivalently, keep solving \eqref{G} with $\xi=\chi=0$ there)
and define $g\df\dot G$.
\par
Note that on $A$, we have $g-\psi_F G=1$, i.e.
\bq
g-\gamma^{-1}e^{-F}G&=&1\eq
\par
It follows that for $r\geq 0$ sufficiently small, say $r\in[0, r_0)$, with some $r_0>0$,
we have
\bq
g-\gamma^{-1}e^{-F_r}G&\geq &1/2\qquad \hbox{ on } A\eq
($r_0$ depends on $h$ through \eqref{G} via a bound on $G$).
\par
The latter inequality can be written as 
\bq
\pa_rf_r+\pa_r\psi_{F_r}&\geq & \f12\qquad \hbox{ on } A\eq
and we deduce that for $r\in[0,r_0)$,
\bq
f_r+\psi_{F_r}&\geq  & 0\qquad \hbox{ on } A\eq
\par
Due to the definition of $\epsilon$ and  $A$, $f+\psi_F$ is bounded below by $\epsilon/2$ on $[x_1,x_0]\setminus A$.
It follows that up to diminishing $r_0$, we can insure that $f_r+\psi_{F_r}\geq 0$ on $[x_1, x_0]$ for all $r\in[0,r_0)$.\par
Up to imposing $G(x_0)=0$, this is the type of perturbations $f_r$ we are to consider.
\par
Note that \eqref{G} can be solved explicitly:
\bqn{jG}
\fo x\in[x_1, x_0],\qquad
G(x)&=& e^{H(x)}\int_{x_1}^x e^{-H(u)}\xi (u)\, du\eqn
where
\bqn{H}
\fo x\in[x_1, x_0],\qquad
H(x)&=&\int_{x_1}^x\chi(u)\, du\eqn
\par
Thus the condition $G(x_0)=0$ writes
\bq
\int_{x_1}^{x_0} e^{-H}\xi \, d\lambda&=&0\eq
namely
\bqn{G0}
\int_{A} e^{-H}\, d\lambda+\int_{B} e^{-H}h\, d\lambda&=&0\eqn
\par
Once this condition is satisfied, we have $f_r\in \cD_{\gamma}$ for $r\in[0,r_0)$.
It leads us to investigate $\cG_2(f_r)$. Let us differentiate this quantity at $r=0$.
First note that
\bq
\pa_r\vert_{r=0}( f_r+\nu_{F_r})&=& g-\nu_F G\eq
\par
If $x\in[x_1,x_0]$ is such that $f(x)+\nu_F(x)=0$, then $x$ does not belong to $A\sqcup B$, so
 $g-\nu_FG=0$.
 Taking into account the definition of $B$,
we obtain by differentiation under the integral 
\bq
\pa_r\vert_{r=0}\cG_2(f_r)&=&\int_B g-\nu_FG\, d\lambda
=\int_B h\,d\lambda\eq
\par
Since $f$ is a global minimizer of $\cG$ on $\cD_{\gamma}$ and that
$\cG(f_r)-\cG(f)=\cG_2(f_r)-\cG_2(f)$, we must have
\bqn{Bh}
\int_B h\,d\lambda&\geq &0\eqn
\par
So we have shown that if $h$ is such that \eqref{G0} is true, then \eqref{Bh} holds.
\par
To finish the proof, it remains to see that this property implies that $\lambda(D)=0$.
\par
We proceed by contradiction, assuming $\lambda(D)>0$. 
Consider $x_2\in (x_1, x_0)$ such 
that
\bqn{DD}
\lambda(D_-)&=&\lambda(D_+)\eqn
with
\bq
D_-\df D\cap [x_1,x_2),\qquad
D_+\df D\cap [x_2, x_0)\eq
\par
Find a bounded and measurable function $h_0$ on $B$
such that
\bqn{h0}
\int_B \exp(-H)h_0\,d\lambda&=&-\int_A \exp(-H)\,d\lambda\eqn
\par
Consider next \bq
h&=&h_0+t\exp(H)\un_{D_-}-t\exp(H)\un_{D_+}\eq
with $t\geq 0$ to be chosen later.
\par
Due to \eqref{DD} and \eqref{h0}, \eqref{G0} holds.
\par
However, as can be seen in  \eqref{H},  $H$ is strictly increasing on $B$.
It follows that
\bq
\fo x'\in D_-,\,\fo x''\in D_+,\qquad H(x')&<&H(x'')\eq
and as a consequence
\bq
\int_Dh\, d\lambda
&=&
\int_D h_0\,d\lambda+t\lt(\int_{D_-}e^H\,d\lambda-\int_{D_+}e^H\,d\lambda\rt)\eq
diverges toward $-\iy$ as $t$ goes to $+\iy$
(the latter assertion follows from $\lambda(D_-)=\lambda(D_+)=\lambda(D)/2>0$).
However, this leads to a contradiction with \eqref{Bh}, so we must have $\lambda(D)=0$, as desired.\wwtbp
\par
We can now come to the 
\proofff{Proof of Theorem \ref{theofs}}
From Proposition \ref{pro0}, we deduce that $\cG_2(f)=0$ and so $\cG(f)=\cG_1(f)$.
Lemma \ref{cG1} tells us that $\cG(f)$ will be minimal if $\z$ is as small as possible.\par\sm
From Proposition \ref{pro0}, we also get that
for  $x\in[\z,x_0]$,
\bqn{fff}
f(x)&\leq & -\f1\gamma\lt(1-\f\alpha{\beta x}\rt)\exp(-F(x))\eqn
inequality which can rewritten under the form
\bq
\f{d}{dx}\exp(F(x))&\leq & -\f1\gamma\f{d}{dx}\lt(x-\f\alpha{\beta }\ln(x)\rt)\eq
(where $d/dx$ corresponds to a weak derivative).
\par
Integrating this bound we obtain
\bq
e^{F(x_0)}-e^{F(\z)}&\leq & -\f1\gamma\lt(x_0-\z-\f\alpha{\beta }\ln(x_0/\z)\rt)\eq
\par
Recalling that 
$F(\z)=0$ and $F(x_0)=\ln(y_0/\gamma)$, we deduce
\bq
\f{y_0}{\gamma}-1&\leq &-\f1\gamma\lt(x_0-\z-\f\alpha{\beta }\ln(x_0/\z)\rt)\eq
i.e.
\bq
1-\gamma-\f\alpha{\beta }\ln(x_0)&\leq &\z-\f\alpha{\beta }\ln(\z)\eq
\par
The r.h.s.\ is a quantity which is increasing with $\z$. Thus we must have
$\z\geq x^*$, where $x^*$ was defined in \eqref{xs}.
\par
The equality $\z=x^*$ is realized if and only if 
\eqref{fff} is an equality (a.e.), namely
\bqn{ffff}
f(x)&= & -\f1\gamma\lt(1-\f\alpha{\beta x}\rt)\exp(-F(x))\eqn
\par
Integrating this equality as before, we get 
\bq
\fo x\in[x^*,x_0],\qquad
e^{F(x)}-1&=& -\f1\gamma\lt(x-x^*-\f\alpha{\beta }\ln(x/x^*)\rt)\eq
\par
Replacing this value of $e^{F(x)}$ in \eqref{ffff},
we get the function announced in \eqref{fs2}.
\wwtbp
\par
In the above arguments also enable to compute the minimal value of $\cG$ on $\cD_{\gamma}$
(which is also the minimal value of $\cK$ on $\cM_{\gamma}$ according to Subsection \ref{rtam}).
\begin{cor}\label{cor}
\bq
\min_{\cD_{\gamma}}\cG&=&\cG(f^*)
=\f\alpha{\beta\gamma}\lt(\ln\lt(\f{\alpha}\beta\rt)-1+\f\beta\alpha-\ln(x_0)\rt)-1\eq
\end{cor}
\prooff
In the proof of Theorem \ref{theofs} we have seen that $\cG(f^*)=\cG_1(f^*)$.
Using the expression given in Lemma~\ref{cG1}, where $\z$ is replaced by $x^*$, we get
\bq
\cG_1(f)&=&\f1\gamma\lt(x^*-\f\alpha\beta\ln(x^*)\rt)+\f\alpha{\beta\gamma}\lt(\ln\lt(\f{\alpha}\beta\rt)-1\rt)\eq
It remains to take into account the characterization \eqref{xs} of $x^*$.
\wwtbp
\par
Remember this value of $\min_{\cD_{\gamma}}\cG$ is only valid under our underlying assumption, otherwise this minimum is simply 0,
as it is attained at the \textit{laissez-faire} policy. 
It should be noted that $\min_{\cD_{\gamma}}\cG$ is decreasing with respect to $\gamma$, as long as our underlying assumption is satisfied (recall \eqref{LF} in the main text).
This observation will be useful in the next subsection.

\subsection{Back to Theorem \ref{Thm: Main}}\label{btt}

Finally we come to the proof of Theorem \ref{Thm: Main}. But first we have to return to a rigorous justification of the restriction to \eqref{abg}, which was only heuristically discussed at the end of
Subsection \ref{txps}. We will also present an extension of Theorem \ref{Thm: Main}.\par\sm
In the setting of Subsection \ref{txps}, consider a function $\varphi\st [\alpha/\beta, x_0]\ri [0,\gamma]$ with $\varphi(\alpha/\beta)<\gamma$, and satisfying (H2), (H3), (H4) and (H5).
Due to the fact that $\varphi$ is right continuous and only jumps upward, this function attains its maximum, say at $x_1\in  [\alpha/\beta, x_0]$.
Define a new function $\wi \varphi$ via
\bq
\fo x\in[\alpha/\beta, x_0],\qquad
\wi\varphi(x)&\df&\lt\{\begin{array}{ll}
\varphi(x_1)&\hbox{, if $x\leq x_1$}\\
\varphi(x)&\hbox{, if $x> x_1$}
\end{array}\rt.\eq
\par
Note that $\wi \varphi$ still satisfies (H2), (H3), (H4) and (H5).
Recall  the functional $\cJ$ defined in \eqref{cJ2}.
\begin{lem}
We have
\bq
\cJ(\wi\varphi)&\leq & \cJ(\varphi)\eq
and the inequality is strict if $\wi\varphi\neq\varphi$.
\end{lem}
\prooff
The argument is similar to the proof of Lemma \ref{l1} and is based on the following computation:
\bq
\int_{\alpha/\beta}^{x_1} L(u,\varphi(u),\varphi'(u))\, du &\geq & 
\f\beta\alpha\int_{\alpha/\beta}^{x_1}  \f{1+\varphi'(u)}{\varphi(u)}-\f{\alpha}{\beta u\varphi(u)}\, du \\
&=&[\ln(\varphi(u)]_{\alpha/\beta}^{x_1}+
\f\beta\alpha\int_{\alpha/\beta}^{x_1}  \f{1}{\varphi(u)}\lt(1-\f{\alpha}{\beta u}\rt)\, du\\
&\geq &[\ln(\varphi(u)]_{\alpha/\beta}^{x_1}+\f\beta{\alpha\varphi(x_1)}\int_{\alpha/\beta}^{x_1} 1-\f{\alpha}{\beta u}\, du\\
&=&\ln(\varphi(x_1)/\varphi(\alpha/\beta))+\f\beta{\alpha\varphi(x_1)}\lt[u-\f{\alpha}{\beta }\ln(u)\rt]_{\alpha/\beta}^{x_1}\\
&\geq &\f\beta{\alpha\varphi(x_1)}\lt[u-\f{\alpha}{\beta }\ln(u)\rt]_{\alpha/\beta}^{x_1}
\eq
\par
If we replace $\varphi$ by $\wi\varphi$ in the above computations, all the inequalities become equalities,
so the last term is in fact equal to
\bq
\int_{\alpha/\beta}^{x_1} L(u,\wi\varphi(u),\wi\varphi'(u))\, du 
\eq
\par
 We have  furthermore 
\bq
\sum_{u\in (\alpha/\beta, x_1]\st \varphi(u)\neq\varphi(u-)} \ln\lt(\f{\varphi(u)}{\varphi(u-)}\rt)&\geq &0
=\sum_{u\in (\alpha/\beta, x_1]\st \wi\varphi(u)\neq\wi\varphi(u-)} \ln\lt(\f{\wi\varphi(u)}{\wi\varphi(u-)}\rt)\eq
and 
the respective contributions of $\varphi$ and $\wi\varphi$ to the costs $\cJ(\varphi)$ and $\cJ(\wi\varphi)$ are the same on $(x_1, x_0]$.
It follows that
\bq
\cJ(\wi\varphi)&\leq & \cJ(\varphi)\eq
\par
The equality holds if in the above computation of the integral \linebreak $\int_{\alpha/\beta}^{x_1} L(u,\varphi(u),\varphi'(u))\, du$, all inequalities are equalities
and we get that $\varphi(x)=\varphi(x_1)$ for a.e.\ $x\in(\alpha/\beta, x_1)$, namely $\wi\varphi=\varphi$.
\wwtbp\par
It follows 
that in the perspective of minimizing $\cJ$, we can restrict our attention to functions $\varphi$ attaining their maximum at $\alpha/\beta$,
i.e.\ we can replace (H1a) by
\begin{itemize}
\item[(H1c):] $\varphi$ is defined on $[\alpha/\beta,x_0]$, takes values in $[0,\gamma]$,
$\varphi(\alpha/\beta)=\max_{[\alpha/\beta,x_0]}\varphi$, $\varphi(x_0)=y_0$ and the left limits of $\varphi$ are positive.
\end{itemize}
\par
To go further toward (H1b), 
let $\varphi$ a function satisfying (H1c), (H2), (H3), (H4) and (H5), denote $\eta\df \varphi(\alpha/\beta)$ and assume that $\eta<\gamma$.
From the development of Subsections \ref{eoJtm} to \ref{cotmoGoD}, we get
that
\bq
\cJ(\varphi)&\geq & \min_{\cD_\eta}\cG
\eq
According to Corollary \ref{cor} (see also its following paragraph), we have
\bq
\min_{\cD_\eta}\cG&>& \min_{\cD_{\gamma}}\cG\eq
\par
This observation ends up the justification of the replacement of (H1a) by (H1b), relatively to the search of a global minimizer of $\cL$.
\par\me
We can now come to the
\proofff{Proof of Theorem \ref{Thm: Main}}
One direct way would be to check that the procedure described in Subsection \ref{txps} transform $b^*$ into $f^*$.
There is even a faster way, as it is sufficient to check that
\bq
\cC(b^*)&=&\cG(f^*)\eq
and this is a consequence of \eqref{I}, on one hand, and of 
Corollary \ref{cor}, on the other hand (recall that $x^*=x(\tau_1)$).\par
This argument shows that $b^*$ is a global minimizer of $\cC$ over $\cB_{\gamma}$. To see that any global minimizer coincides with $b^*$
on the time interval $[0,\tau_1]$, take into account Remark \ref{inv1}: since $f^*$ is Lipschitzian, there is only one policy leading to $f^*$
(defined on $[\alpha/\beta, x_0]$).
\wwtbp
\par
Let us end this subsection with several observations.
\begin{rem}
Corollary \ref{cor} provides in fact a quantitative formulation of our underlying assumption, since it does correspond to 
$\min_{\cD_{\gamma}}\cG>0$ and we get 
\bq
\gamma&<&\f\alpha{\beta}\lt(\ln\lt(\f{\alpha}\beta\rt)-1+\f\beta\alpha-\ln(x_0)\rt)\eq
(in particular the r.h.s.\ must be positive). We recover \eqref{LF} in the main text.
\end{rem}
\par
\begin{rem}\label{r2}
Consider a cost functional of the form \eqref{wiC}, where $F$ coincides with $(\cdot)_+$ on $\RR_+$ and is non-negative on $(-\iy,0)$.
Then we get that $\wi \cC\geq \cC$. Note nevertheless that $\wi\cC(b^*)=\cC(b^*)$. It follows that $b^*$ is also a global minimizer of $\wi\cC$.
In particular if $F$ is positive on $(-\iy,0)$, then $b^*$ is the unique minimizer of $\wi \cC$.
\end{rem}
\begin{rem}
Our extension of the optimization problem to measure spaces suggests that the S.I.R.\  ODE\ \eqref{ODE3} could itself be generalized into
\bqn{SIR2}
\lt\{
\begin{array}{rcl}
dx&=& -xy\, dB\\[2mm]
dy&=& xy\,dB-\alpha y\,dt
\end{array}
\rt.\eqn
where $B$ is a Radon signed measure on $\RR_+$ (in \eqref{ODE3}, it is given by $B([0,t])=\int_0^t b(s)\, ds$, for all $t\geq 0$).
Equation \eqref{SIR2} is to be understood in the Stieltjes sense: for any $t\geq 0$,
\bq
\lt\{
\begin{array}{rcl}
x(t)&=& x(0)-\int_{[0,t]}x(s)y(s) \, dB(s)\\[2mm]
y(t)&=&y(0)+ \int_{[0,t]}x(s)y(s)\,dB(s)-\int_{[0,t]}\alpha y(s)\,ds
\end{array}
\rt.\eq
(where $x$ and $y$ are themselves only right-continuous with left limits, in fact they should be seen as repartition functions of measures).
It would be modeling very erratic policies and Theorem \ref{theofs}
would imply that even among them, $b^*$ is a minimizer of the extension of $\cC$ similar to $\cJ$ (as one would guess).
\end{rem}
\par

\end{document}